%% file: main.tex
\documentclass[review]{elsarticle}
\usepackage[utf8]{inputenc}

\usepackage[hyphens]{url}
\usepackage{amsmath,amsfonts,amssymb,amscd,amsthm}
\usepackage{mathtools}
\theoremstyle{plain}

\theoremstyle{definition}

\theoremstyle{remark}

\usepackage{changes}

\usepackage[noend,linesnumbered,ruled,noline]{algorithm2e}

\usepackage[colorlinks = true,
linkcolor = blue,
urlcolor  = blue,
citecolor = blue,
anchorcolor = blue]{hyperref}

\usepackage{tikz}
\usepackage{nicematrix}

\usepackage{bm}

\usepackage[margin=2.5cm]{geometry}

\usepackage[T1]{fontenc}
\usepackage[utf8]{inputenc}
\usepackage[english]{babel}
\usepackage[autostyle=true]{csquotes} 
\usepackage{todonotes}

\usepackage{graphicx}
\graphicspath{{figures/}}   
\usepackage{tikz}

\usepackage{caption}
\usepackage{subcaption}

\usepackage{multirow}
\usepackage{booktabs}

\title{Variational Bayesian Approximation of Inverse Problems using Sparse  Precision Matrices}

\input{defs.tex}

\begin{document}

\begin{frontmatter}
\author[1,3]{Jan Povala\corref{cor1}%
\fnref{fn1}}
\ead{jan.povala@gmail.com}

\author[2]{Ieva Kazlauskaite\corref{cor1}\fnref{fn1}}
\ead{ik394@cam.ac.uk}

\author[2,3]{Eky Febrianto}
\author[2,3]{Fehmi Cirak}
\author[2,3]{Mark Girolami}

\cortext[cor1]{Corresponding authors}
\fntext[fn1]{Equal contribution.}

\address[1]{Department of Mathematics, Imperial College London, London, SW7 2AZ, UK}
\address[2]{Department of Engineering, University of Cambridge, Cambridge, CB2 1PZ, UK}
\address[3]{The Alan Turing Institute, London, NW1 2DB, UK}

\begin{abstract}
Inverse problems involving partial differential equations (PDEs) are widely used in science and engineering. Although such problems are generally ill-posed, different regularisation approaches have been developed to ameliorate this problem. Among them is the Bayesian formulation, where a prior probability measure is placed on the quantity of interest. The resulting posterior probability measure is usually analytically intractable. The Markov Chain Monte Carlo (MCMC) method has been the go-to method for sampling from those posterior measures. MCMC is computationally infeasible for large-scale problems that arise in engineering practice.  Lately, Variational Bayes (VB) has been recognised as a more computationally tractable method for Bayesian inference, approximating a Bayesian posterior distribution with a simpler trial distribution by solving an optimisation problem. In this work, we argue, through an empirical assessment, that VB methods are a flexible and efficient alternative to MCMC for this class of problems. We propose a natural choice of a family of Gaussian trial distributions parametrised by precision matrices, thus taking advantage of the inherent sparsity of the inverse problem encoded in its finite element discretisation. We utilise stochastic optimisation to efficiently estimate the variational objective and assess not only the error in the solution mean but also the ability to quantify the uncertainty of the estimate. We test this on PDEs based on the Poisson equation in 1D and 2D. A Tensorflow implementation is made publicly available on GitHub.
\end{abstract}

\end{frontmatter}

\section{Introduction}

The increased availability of measurements from engineering systems allows for the development of new and the improvement of existing computational models, which are usually formulated as partial differential equations. 
Inferring model parameters from observations of the physical system is termed the \emph{inverse problem}~\citep{tarantolaInverseProblemTheory2005, kaipioStatisticalComputationalInverse2005, stuartInverseProblemsBayesian2010}. 
In this work, we consider the inverse problem where the quantities of interest (for example, some material properties) and the observations (e.g., the displacement field)  are related through elliptic PDEs. 
Most inverse problems are non-linear and ill-posed, meaning that the existence, uniqueness, and/or stability (continuous dependence on the parameters) of the solution are violated \citep{stuartInverseProblemsBayesian2010, tarantolaInverseProblemTheory2005, kaipioStatisticalComputationalInverse2005}. 
These issues are often alleviated through some regularisation, like Tikhonov regularisation~\citep{tikhonovSolutionsIllposedProblems1977}, that imposes assumptions on the regularity of the solution. Alternatively, the specification of the prior in the Bayesian formulation of inverse problems provides a natural choice for regularisation, and any given regularisation can be interpreted as a specific choice of priors in the Bayesian setting~\citep{bishopPatternRecognitionMachine2006}. Furthermore, the Bayesian formulation provides not only a qualitative but also a quantitative estimate of both epistemic and aleatoric uncertainty in the solution. In particular, the mean of the posterior probability distribution corresponds to the point estimate of the solution while the credible intervals capture the range of the parameters consistent with the observed measurements and prior assumptions. For these reasons, Bayesian methods have gained popularity in computational mechanics for experimental design and inverse problems with uncertainty quantification; see, e.g., the recent papers by~\citet{abdulleProbabilisticFiniteElement2021, panditaSurrogatebasedSequentialBayesian2021, pyrialakosNeuralNetworkaidedBayesian2021, niProbabilisticModelUpdating2021, sabaterBayesianApproachQuantile2021, huangSequentialSparseBayesian2021, ibrahimbegovicReducedModelMacroscale2020, tarakanovOptimalBayesianExperimental2020, michelenstroferEnforcingBoundaryConditions2020, carlonNesterovaidedStochasticGradient2020, wuBayesianInferenceNonlinear2020, uribeBayesianInferenceRandom2020, rizziBayesianModelingInconsistent2019,arnstIdentificationSamplingBayesian2019, beckFastBayesianExperimental2018, betzBayesianInferenceSubset2018, chenHessianbasedAdaptiveSparse2017, asaadiComputationalFrameworkBayesian2017, huangBayesianSystemIdentification2017, karathanasopoulosBayesianIdentificationTendon2017, babuskaBayesianInferenceModel2016, girolamiStatisticalFiniteElement2021}.

The Bayesian formulation of inverse problems is also the focal point of probabilistic machine learning, and in recent years significant progress has been made in adapting and scaling machine learning approaches to complex large-scale problems~\citep{luSurpassingHumanLevelFace2015,solinPIVOProbabilisticInertialVisual2018}. One of the leading models for Bayesian inverse problems are Gaussian processes (GPs) which define probability distributions over functions and allow for incorporating observed data to obtain posterior distributions. Given that most posterior distributions in Bayesian inference are analytically intractable, approximation methods need to be resorted to. Two classical approximation schemes are the Markov Chain Monte Carlo (MCMC) and the Laplace approximation. The MCMC algorithm proceeds by creating a Markov Chain whose stationary distribution is the desired posterior distribution. Although MCMC provides asymptotic convergence in distribution, devising an efficient, finite-time sampling scheme is challenging, especially in higher dimensions \citep{gelmanBayesianDataAnalysis2013}. 
Application-specific techniques such as parameter space reduction and state space reduction have been proposed in the literature to help scale up MCMC methods, but these low-rank approximations are not specific to MCMC methods only~\citep{cuiScalablePosteriorApproximations2016}. 
Due to the asymptotic correctness of MCMC, we use it as a benchmark for the experimental studies in this paper. 
Meanwhile, the Laplace approximation finds a Gaussian density centred around the mode of the true posterior, utilising the negative Hessian of the unnormalised posterior log-density \citep{bishopPatternRecognitionMachine2006}.
The Hessian is a large dense matrix, where forming each column requires multiple PDE solves; to make such calculations feasible, low-rank approximations are typically used~\citep{villaHIPPYlibExtensibleSoftware2021, bui-thanhComputationalFrameworkInfiniteDimensional2013}. Evidently, the Laplace approximation is not suitable for multi-modal posterior distributions due to the uni-modality of the Gaussian distribution. 

\subsection{Related work}
%
In recent years, advances in variational Bayes (VB) methods have allowed for Bayesian inference to be successfully applied to large data sets. Variational Bayes translates a sampling problem that arises from applying the Bayes rule into an optimisation problem~\citep{jordanIntroductionVariationalMethods1999,bleiVariationalInferenceReview2017, jordanGraphicalModelsExponential2007}. The method finds a solution that minimises the Kullback-Leibler (KL) divergence between the true posterior distribution and a trial distribution from a chosen family of distributions, for instance, multivariate Gaussian distributions with a specific covariance structure.  The strong appeal of VB is that one can explicitly choose the complexity of the trial distribution, i.e., its number of free parameters, such that the resulting optimisation problem is computationally tractable, and the approximate posterior adequately captures important aspects of the true posterior. 

Further scalability of VB methods is due to advancements in sparse approximations and approximate inference.
For instance, sparse GP methods such as Nystr\"om approximation or fully independent training conditional method (FITC) rely on lower-dimensional representations that are defined by a smaller set of so-called inducing points to represent the full GP~\citep{williamsUsingNystromMethod2001, csatoSparseOnlineGaussian2002, seegerFastForwardSelection2003, quinonero-candelaUnifyingViewSparse2005, snelsonSparseGaussianProcesses2006, titsiasVariationalLearningInducing2009, titsiasVariationalModelSelection2008}. Using this approximation for a data set of size $N$, algorithmic complexity is reduced from $\mathcal{O}(N^3)$ to $\mathcal{O}(NM^2)$, while storage demands go down from $\mathcal{O}(N^2)$ to $\mathcal{O}(NM)$, where $M$ is a user selected number of inducing variables. To widen the applicability of VB to large datasets and non-conjugate models (combinations of prior distributions and likelihoods that do not result in a closed-form solution), \emph{stochastic variational inference} (SVI) was proposed~\citep{hensmanFastVariationalInference2012, hoffmanStochasticVariationalInference2013, hensmanGaussianProcessesBig2013}. Sub-sampling the original data and Monte Carlo estimation of the optimisation objective and its gradients, allows for calibrating complex models using large amounts of data. Multiple further extensions to the sparse SVI framework were proposed, leveraging the Hilbert space formulation of VB~\citep{chengVariationalInferenceGaussian2017}, introducing parametric approximations~\citep{jankowiakParametricGaussianProcess2020}, applying the Lanczos algorithm to efficiently factorise the covariance matrix~\citep{pleissConstanttimePredictiveDistributions2018}, transforming to an orthogonal basis~\citep{salimbeniOrthogonallyDecoupledVariational2018, shiSparseOrthogonalVariational2020}, and adapting to compositional models~\citep{salimbeniDoublyStochasticVariational2017}. 

The choice of prior is a central task in designing Bayesian models. If the prior is obtained from a domain expert, it is not necessarily less valuable than the data itself; one way of thinking about a prior is by considering how many observations one would be prepared to trade for a prior from an expert -- if the expert is very knowledgeable, then one might be prepared to exchange a large part of a dataset to get access to that prior. Translating the expert knowledge into a prior probability distribution is a challenging task, and due to practical considerations, certain choices of priors are preferred for their simplicity and analytic tractability. When inferring values of parameters over a spatial domain, as is typically the case in finite elements, GP priors offer a natural way to incorporate the information about the smoothness and other known properties of the solution.
We note that while other Bayesian models, such as Bayesian neural networks are gaining interest, it is very difficult to impose functional priors in such models, challenging the effective use of expert knowledge and leading to unrealistic uncertainty estimates~\citep{sunFunctionalVariationalBayesian2019, burtUnderstandingVariationalInference2021}.

\subsection{Contributions}
%
In this work, we advocate for the use of GP priors with stochastic variational inference as a principled and efficient way to solve the inverse problems arising in computational mechanics. We show, through an extensive empirical study, that variational Bayes methods provide a flexible and efficient alternative to MCMC methods in the context of Bayesian inverse problems based on elliptic PDEs while retaining the ability to quantify uncertainty. While similar directions have been explored in previous work, the focus there is on specific applications, such as parameter estimation problems in models of contamination~\citep{tsilifisComputationallyEfficientVariational2016} or proof-of-concept on particular 1D inverse problems~\citep{barajas-solanoApproximateBayesianModel2019}. 

We extend the previous works in multiple aspects, focusing on improving the utility of VB in inverse problems arising from elliptic PDEs and providing a thorough discussion of the empirical results that can be used by practitioners to guide their use of VB in applications.
Specifically, we argue that the efficiency of the VB algorithms for PDE based inverse problems can be improved by taking into account the structure of the problem, as encoded in the FEM discretisation of the PDE. 
Motivated by previous uses of precision matrices as a way of describing conditional independence~\citep{tanGaussianVariationalApproximation2018,durrandeBandedMatrixOperators2019}, we leverage the sparse structure of the problems to impose conditional independence in the approximating posterior distribution. 
This choice of parametrisation results in sparse matrices, which improve the computational and the memory cost of the resulting algorithms. 
Such parametrisation, combined with stochastic optimisation techniques, allows the method to be scaled up to large problems on 2D domains.
Through extensive empirical comparisons, we demonstrate that VB provides high-quality point estimates and uncertainty quantification comparable to the estimates attained by MCMC algorithms but with significant computational gains.
Finally, we describe how the proposed framework can be seamlessly combined with existing solvers and optimisation algorithms in the finite element implementations. 

The main concern related to VB in statistics stems from the fact that it is constrained by the chosen family of trial distributions, which may not approximate the true posterior distribution well. If the choice of the trial distributions is too restrictive, the estimate of the posterior mean is biased while the uncertainty may be underestimated~\citep{mackayInformationTheoryInference2003, wangInadequacyIntervalEstimates2005, turnerTwoProblemsVariational2011}. Furthermore, as noted in previous work, the commonly used mean-field factorisation of the trial distributions does not come with general guarantees on accuracy~\citep{giordanoCovariancesRobustnessVariational2018}. However, VB has been demonstrated to work well in practice in a variety of settings~\citep{kingmaAutoencodingVariationalBayes2014, damianouVariationalInferenceLatent2016, bleiVariationalInferenceReview2017, zhangAdvancesVariationalInference2019}. Recent work on VB has provided some tools for assessing the robustness of the VB estimates~\citep{giordanoCovariancesRobustnessVariational2018} . 
\subsection{Overview}
%
The rest of the paper is structured as follows. In Sec.~\ref{sec:formulation}, we define Bayesian inverse problems and detail some inference challenges related to their ill-posedness. In Sec.~\ref{sec:main_vb_section}, we give a presentation of the variational Bayes framework, with strong focus on sparse parametrisation resulting from conditional independence. We give details of the experiments and the evaluation criteria, and discuss obtained results for each experiment in Sec.~\ref{sec:examples}. Lastly, Sec.~\ref{sec:conclusion} concludes the paper and discusses some promising directions for future work.

\section{Bayesian Formulation of Inverse Problems \label{sec:formulation}}
In this section, we review the Bayesian formulation of inverse problems by closely following~\citet{stuartInverseProblemsBayesian2010}.
%
\subsection{Forward map and observation model}
We are interested in finding $\kappa \in \mathcal{K}$, an input to a model, given $\y \in \mathcal{Y}$, a noisy observation of the solution of the model, where $\mathcal{K}, \mathcal{Y}$ are Banach spaces\footnote{Respective norms for Banach spaces $\mathcal{K}$, $\mathcal{Y}$ are $\Vert\cdot\Vert_{\mathcal{K}}$ and $\Vert\cdot\Vert_{\mathcal{Y}}$.}. The mapping is given by
\begin{equation} \label{eq:fwd-problem-general}
    \y = \mathcal{G}(\kappa) + \eta,
\end{equation}
where $\mathcal{G}: \mathcal{K} \rightarrow \mathcal{Y}$, $\eta \in \mathcal{Y}$ is additive observational noise. We focus on problems where $\mathcal{G}$ maps solutions of elliptic partial differential equations with input $\kappa \in \mathcal{K}$ into the observation space $\mathcal{Y}$. For a suitable Hilbert space $\mathcal{U}$, which we make concrete later, let $\mathcal{A} \colon \mathcal{K} \rightarrow \mathcal{U}$ be a possibly non-linear solution operator of the PDE. For a particular $\kappa \in \mathcal{K}$, the solution $u \in \mathcal{U}$ is
\begin{equation}
    u = \mathcal{A}(\kappa).
\end{equation}
To obtain observations $\y$, we define a projection operator $\mathcal{P} \colon \mathcal{U} \rightarrow \mathcal{Y}$. Consequently,~\eqref{eq:fwd-problem-general} can be written out in full as 
\begin{equation}
    \y = \mathcal{P}(\mathcal{A}(\kappa)) + \eta.
\end{equation}

\subsection{Inference}
We solve the inverse problem~\eqref{eq:fwd-problem-general} for $\kappa$ by finding $\kappa$ such that the data misfit, $\Vert \y - \mathcal{G}(\kappa) \Vert_{\mathcal{Y}}$, is minimised. 
As already mentioned in the introduction, this is typically an ill-posed problem: there may be no solution, it may not be unique, there may exist a dimensionality mismatch between the observations and the quantity being inferred, and it may depend sensitively on $y$. To proceed, we choose the Bayesian framework for regularising the problem to make it amenable to analysis and practical implementation. We describe our prior knowledge about $\kappa$ in terms of a prior probability measure $\mu_0$ on the subspace of $\mathcal{K}$ and use Bayes' formula to calculate the posterior probability measure, $\mu^{\y}$, for $\kappa$ given $\y$. The relationship between the posterior and prior is expressed as
\begin{equation}
    \frac{\mathrm{d}\mu^\y}{\mathrm{d}\mu_0}(\kappa) = \frac{1}{Z(\y)}\exp(-\Phi(\kappa; \y)),
    \label{eq:radon_nikodym_derivative_infinite_dim}
\end{equation}
where $\frac{\mathrm{d}\mu^\y}{\mathrm{d}\mu_0}$ is the Radon-Nikodym derivative of $\mu^\y$ with respect to $\mu_0$, and $\Phi$ is the potential function which is determined by the forward problem~\eqref{eq:fwd-problem-general}, specifically $\mathcal{G}$ and $\eta$. To ensure that $\mu^{\y}$ is a valid probability measure, we have  $Z(y) = \int_{\mathcal{K}} \exp(-\Phi(\kappa; \y)) \mathrm{d} \mu_0(\kappa)$.

From here on, we assume that $(\mathcal{Y}, \Vert \cdot \Vert_{\mathcal{Y}}) = (\RR^{n_{\y}}, \Vert \cdot \Vert)$, where $\Vert \cdot \Vert$ is the Euclidean norm, and we treat data $\y$ and $\eta$ as vectors, i.e. $\vy$ and $\veta$. We specify the additive noise vector $\veta$ as the zero-mean Gaussian with covariance matrix $\boldsymbol{\Gamma}$, such that 
$$
\veta \sim \mathcal{N}(\mathbf{0}, \mathbf{\Gamma} = \sigma_{\y}^2\mathbf{I}),
$$
where $\sigma_y$ is the standard deviation of the measurement noise and $\mathbf{I}$ is the identity matrix. We can write $\Phi$ conveniently as
\begin{equation}
    \label{eq:misfit-potential-function}
    \Phi(\kappa; \vy) = \frac{1}{2} \Vert \mathcal{G}(\kappa) -  \vy\Vert_{\boldsymbol{\Gamma}}^2,
\end{equation}
where $\Vert \cdot \Vert_{\boldsymbol{\Gamma}}$ is the norm induced by the weighted inner product.\footnote{For any self-adjoint positive operator $\mathcal{T}$, the weighted inner product is $\langle \cdot, \cdot \rangle_{\mathcal{T}} = \langle \mathcal{T}^{-1/2}\cdot, \mathcal{T}^{-1/2} \cdot \rangle$, and the induced norm is $\Vert \cdot \Vert_{\mathcal{T}} = \Vert \mathcal{T}^{-1/2} \cdot \Vert$.}

We restrict the space of solutions $\mathcal{K}$ to be a Hilbert space and place a Gaussian prior measure on $\kappa$ with mean $m$ and covariance operator $\mathcal{C}_\kappa$ such that
\begin{equation}
\mu_0(\kappa) \sim \mathcal{N}(m, \mathcal{C}_\kappa).
\end{equation}
%
For detailed assumptions on $\mu_0$, $\mathcal{G}$, and $\eta$ that are required for deriving the posterior probability measure, we refer the reader to \citet[Sec.~2.4]{stuartInverseProblemsBayesian2010}.

\subsubsection{Algorithms}

The objective is to find the posterior measure $\mu^\y$ conditioned on the observations, as dictated by Bayes's rule. The forward map~\eqref{eq:fwd-problem-general} and the respective functions  must be discretised.  In Bayesian inference there are two possible approaches for discretisation:~1)~apply the Bayesian methodology first, discretise afterwards, or 2)~discretise first, then apply the Bayesian methodology \citep{stuartInverseProblemsBayesian2010}. 

The first approach develops the solution of the inference problem in the function space before discretising it. A widely used algorithm of this form is the pre-conditioned Crank-Nicholson (pCN) MCMC scheme, where proposals are based on the prior measure $\mu_0$ and the current state of the Markov chain. The pCN method is a standard choice for high-dimensional sampling problems, as its implementation is well-defined and is invariant to mesh refinement~\citep{cotterMCMCMethodsFunctions2013, pinskiAlgorithmsKullbackLeibler2015}. Since we will use this algorithm as one of the baselines, a summary of the algorithm is provided in~\ref{sec:pcn_algo_section}. Recently, infinite-dimensional MCMC schemes that leverage the geometry of the posterior to improve the efficiency have been proposed, see~\citet{beskosGeometricMCMCInfinitedimensional2017}. Other than MCMC schemes, some variational Bayes formulations in function space have been proposed (for example,~\citet{minhInfinitedimensionalLogDeterminantDivergences2017,burtUnderstandingVariationalInference2021}), though currently they do not offer a viable computational alternative to the finite-dimensional formulation of variational inference. 

The second approach proceeds by first discretising the problem and then deriving the inference method. This approach forms the basis of almost all inference procedures developed in engineering: MCMC algorithms such as Metropolis-Hastings~\citep{metropolisEquationStateCalculations1953, hastingsMonteCarloSampling1970} or Hamiltonian Monte Carlo (HMC)~\citep{duaneHybridMonteCarlo1987},  the Laplace approximation, or variational Bayes~\citep{jordanVariationalFormulationFokkerplanck1998, jordanIntroductionVariationalMethods1999}  are used to approximate the posterior.  In the discretised formulation, HMC has achieved recognition as the \textit{gold standard} for its good convergence properties, favourable performance on high-dimensional and poorly conditioned problems, and universality of implementation that enables its generic use in many applications through probabilistic programming languages (\emph{e.g.}, Stan~\citep{carpenterStanProbabilisticProgramming2017}). Therefore, along with the pCN scheme mentioned above, our baseline for inference methods includes the HMC method, and we provide a summary of the HMC scheme in Sec.~\ref{sec:hmc_algo_section}.

For the rest of the exposition in this paper, we will focus on algorithms in the finite-dimensional case, where we discretise $\kappa$ to yield a vector $\kbf$. In finite dimensions, probability densities with respect to the Lebesgue measure can be defined, thus leading to a more familiar form of the Bayes's rule:
\begin{equation}
  p(\kbf \mid \vy) 
    = \frac{p(\vy \mid \kbf)~p(\kbf)}{p(\vy)}  \propto  p(\vy \mid \kbf)~p(\kbf) , \label{eq:model}
\end{equation} 
where $p(\kbf \mid \vy )$  is the posterior density, $p(\vy \mid \kbf)$ is the likelihood of the observed data $\vy$ for a given discretised~$\kbf$ and is determined by the discretised forward problem~\eqref{eq:fwd-problem-general} and noise $\veta$. The prior density for~$\boldsymbol{\kappa}$, which itself may depend on some (hyper-) parameters $\psi$, is denoted by $p(\boldsymbol{\kappa})$. Next two sections focus on discussing $p(\vy \mid \kbf)$ and $p(\kbf)$, respectively.

\subsection{Poisson Equation and likelihood}
%
Let us consider a specific forward problem where $u$ is the solution to the Poisson problem:
\begin{equation}
\begin{aligned}
  \label{eq:poisson-equation-2d}
  -\nabla &\cdot (\exp(\kappa( \vx )) \nabla u( \vx )) = f( \vx ),
\end{aligned}
\end{equation}
where $\vx \in \Omega \subset \RR^d$, with $d\in \{1, 2, 3\}$, $\kappa(\vx)\in \mathbb{R}$ is the log-diffusion coefficient, $u(\vx) \in \mathbb{R}$ is the unknown, and $f(\vx) \in \mathbb{R}$ is a deterministic forcing term. The boundary conditions have been omitted for brevity. We are given $n_\y$ noisy observations $\vy \in \RR^{n_y}$ of the solution~$u$ at a finite set of points, ~$\{ \vx_i \}_{i=1}^{n_y}$. The observation points are collected in the matrix~$\X \in \RR^{n_y \times d}$. Although this PDE is linear in $u$ for a given $\kappa$, the methodology in this paper applies to non-linear cases and can be extended for time-dependent cases such as the inverse problem of inferring initial conditions of a system given observations of the system at a later time.

%
We discretise the weak form of the Poisson problem~\eqref{eq:poisson-equation-2d} with a standard finite element approach. Specifically, the domain of interest $\Omega$ is subdivided into a set $\{\omega_e\}_{e=1}^{n_e}$ of non-overlapping elements of size $h = \max_e \text{diam}(\omega_e)$ such that:
\begin{equation}
\Omega = \bigcup_{e=1}^{n_e} \omega_e \, .
\end{equation}
The unknown field $u(\vx)$ is approximated with Lagrange basis functions $\phi_i(\vx)$ and the respective nodal coefficients $\uu = (u_1, \dots, u_{n_u})^\top$ of the
$n_u$ non-Dirichlet boundary mesh nodes by 
\begin{equation}
u_h(\vx) = \sum_{i=1}^{n_u} \phi_i(\vx) u_i \, . 
\end{equation} 
The discretisation of the weak form of the Poisson equation yields the linear system of equations
\begin{equation}
\Abf(\kbf) \uu = \f \, ,
\label{eq:Akappau}
\end{equation} 
where $\Abf(\kbf) \in \mathbb{R}^{n_u \times n_u}$ is the stiffness matrix, $\kbf \in \RR^{n_{\kbf}}$ is the vector of log-diffusion coefficients, $\f \in \mathbb{R}^{n_u}$ is the nodal source vector. 
The stiffness matrix of an element with label~$e$ is given by
\begin{equation}
A^e_{ij} (\kappa_e) = \int_{\omega_e} \exp(\kappa_e) \frac{\partial \phi_i(\vx)}{\partial \vx} \cdot \frac{\partial \phi_j(\vx)}{\partial \vx} \mathrm{d}\vx \, ,
\end{equation} 
where the log-diffusion coefficient $\kappa_e$ of the element is assumed to be \emph{constant} within the element. The source vector is discretised as:
\begin{equation}
f_i = \int_{\Omega} f(\vx) \phi_i(\vx) \mathrm{d}x \, .
\end{equation} 

Hence, according to the observation model~\eqref{eq:misfit-potential-function} the likelihood is given by 
\begin{equation}
	  p(\vy \mid \kbf) = p(\vy \mid \uu({\kbf})) = \mathcal{N}(\mathbf{P} \Abf (\kbf)^{-1} \f, \sigma_y^2 \mathbf{I}) \, ,
\end{equation}
where the matrix~$\mathbf{P}$ represents the discretisation of the observation operator~$\mathcal{P}$. 

Then the mapping from the coefficients $\kbf$ to the solution $\uu$ is $\uu(\kbf) = \Abf (\kbf)^{-1} \f$. The marginal distribution of $\uu$ is given by:
\begin{equation}
\begin{aligned}
    p(\uu) &= \int p(\uu \mid \kbf) p(\kbf) \mathrm{d}\kbf  \, ,
\end{aligned}
\end{equation} 
where $p(\uu \mid \kbf)$ is deterministic as defined in~\eqref{eq:Akappau} but $\kbf$ appears in it non-linearly, implying that the inference is not analytically tractable.

Throughout the experiments in the later sections, we either set Dirichlet (essential) boundary conditions everywhere (for example $u(\vx) = 0 \; \text{on} \; \partial\Omega$), or assume Neumann (natural) boundary conditions on parts of the boundary. The choice will be made explicit in each experiment. To compute the likelihood, we solve the Poisson problem~\eqref{eq:poisson-equation-2d} for $u(\vx)$ using the finite element method~(FEM).

\subsection{Prior} \label{sec:background-prior-discussion}
%
As discussed above, we place a Gaussian measure on $\kappa$, $\mu_0(\kappa) \sim \mathcal{N}(m, \mathcal{C}_{\kappa})$. Properties of samples from the measure depend on mean $m$ and on the spectral properties of the covariance operator $\mathcal{C}_\kappa$. We restrict the space of prior functions to $L^2(\Omega, \RR)$. Then, operator $\mathcal{C}_\kappa$ can be constructed from the covariance function, $k(\vx, \vx^\prime) = \E \big[ \big(\kappa(\vx) - m(\vx)\big) \big(\kappa(\vx^\prime) - m(\vx^\prime)\big)\big]$ as:
\begin{equation}
    (\mathcal{C}_\kappa \gamma) (\vx) = \int_{\Omega} k(\vx, \vx^\prime) \gamma(\vx^\prime) \mathrm{d}\vx^\prime,
\end{equation}
for any $\gamma \in L^2(\Omega, \RR)$. This formulation is what is commonly referred to as a Gaussian process (GP) with mean function $m(\cdot)$, which we assume to be zero, and covariance function $k(\cdot, \cdot)$ such that 
\begin{equation}
    \kappa \sim \mathcal{GP}\big(m(\cdot), k(\cdot, \cdot)\big).
\end{equation}

Even though the process is infinite-dimensional, an instantiation of the process is finite and reduces to a multivariate Gaussian distribution by definition. The covariance function is typically parametrised by a set of hyperparameters $\psi$.  One popular option, which satisfies assumptions about $\mu_0$ as per~\citet{stuartInverseProblemsBayesian2010}, is the squared exponential kernel (also called the exponentiated quadratic or the radial basis function (RBF) kernel): 
\begin{equation}
    k_{\text{SE}}(\vx, \vx^\prime) = \sigma^2_\kappa \exp \left(-\frac{r^2}{2 \ell_\kappa^2} \right), \label{eq:SEkernel}
\end{equation} 
where $r =\Vert \vx - \vx^\prime \Vert_2$ is the Euclidean distance between the inputs. It depends on two hyper-parameters $\psi = \{\sigma_\kappa, \ell_\kappa\}$, the scaling parameter $\sigma_\kappa$, and the length-scale $\ell_\kappa$. Note that, $k_{\text{SE}}(\cdot, \cdot)$ is an infinitely smooth function, which implies that so is $\kappa(\cdot)$. The RBF kernel imposes smoothness and stationarity assumptions on the solution; in addition, such choice of kernel offers a way to regularise the resulting optimisation problem. However, depending on the expert knowledge of the true solution, other kernels may be used to impose other assumptions such as periodicity.

Both conditioning and marginalisation of the GP can be done in closed form. In particular, consider the joint model of the values $\kbf$ at training locations $\X$ and the unknown test values $\kbf^*$ at test locations $\X^*$:
\begin{gather}
\begin{bmatrix} \kbf \\ \kbf^* \end{bmatrix}  
\sim 
\mathcal{N}\left(\mathbf{0}, 
\begin{bmatrix} \K_{\psi}(\X, \X) & \K_{\psi}(\X, \X^*) \\ \K_{\psi}(\X^*, \X) & \K_{\psi}(\X^*, \X^*)   \end{bmatrix} 
        \right),
        \label{eq:preliminaries:joint}
\end{gather}
where $\K_{\psi}(\X, \X^*)$ is the matrix resulting from evaluating $k(\cdot, \cdot)$ at all pairs of training and test points. The conditional distribution of the function values $\kbf^*$ given the values $\kbf$ at $\X$ is:
\begin{equation} \label{eq:gp_posterior}
    \kbf^* \mid \kbf \sim \mathcal{N}\left( \tilde{\kbf}^*, \: \tilde{\K}  \right),
\end{equation}
where
\begin{equation}
    \begin{aligned}
    \tilde{\kbf}^* &=  \K\left(\X^{*}, \X \right)\left[\K(\X, \X)\right]^{-1} \kbf \\
    \tilde{\K} &= \K\left(\X^{*}, \X^{*}\right)- \K\left(\X^{*}, \X\right)\left[\K(\X, \X)\right]^{-1} \K\left(\X, \X^{*}\right).
\end{aligned}
\end{equation}
The marginal distribution can be recovered by finding the relevant part of the covariance matrix; for example, the marginal of $\kbf$ given $\X$ is $\kbf \sim \mathcal{N}\left( \boldsymbol{0}, \K_{\psi}(\X, \X) \right)$.

In this work, we place a zero-mean Gaussian process prior on $\kappa(\vx)$ and assume the squared exponential kernel with length-scale $\ell_{\kappa}$ and fixed variance $\sigma_{\kappa}^2=1$. As mentioned in the previous section, we assume that $\kappa(\vx)$ is constant on each element of the mesh (we use the same mesh as for discretising $u(\vx)$ and $f(\vx)$). We place the prior on $\kbf$ so that the centroids of the elements are the training points of the GP:
\begin{equation}
    p(\kbf) = \mathcal{N}(\boldsymbol{0}, \K_{\psi}(\X, \X)).
\end{equation}

\section{Variational Bayes Approximation}\label{sec:main_vb_section}

\subsection{Variational Bayes}

We assume that any hyper-parameters $\psi$ of the prior are fixed, and are only interested in the posterior distribution of $\kbf$. 
The variational approach proceeds by approximating the true posterior~$ p(\kbf \mid \vy) $ according to~\eqref{eq:model} with a trial density $q(\kbf)$, which is the minimiser of the discrepancy between a chosen family of trial densities $\mathcal{D}_q$ and the true posterior distribution $p(\kbf | \vy)$ \citep{jordanIntroductionVariationalMethods1999, jordanGraphicalModelsExponential2007}. A typical choice for the measure of discrepancy between distributions is the Kullback-Leibler (KL) divergence (which due to the lack of symmetry is not a metric). To find the approximate posterior distribution we have:
\begin{equation}
  q^{*}(\kbf) = \underset{q(\kbf) \in \mathcal{D}_q}{\argmin} \, \kl{q(\kbf)}{p(\kbf \mid \vy)}.
\end{equation}
Expanding the KL divergence term we obtain
\begin{equation}
\begin{aligned}
  \kl{q(\kbf)}{p(\kbf \mid \vy)} 
  & = \int q(\kbf) \log \frac{q(\kbf)}{p(\kbf \mid \vy)} \mathrm{d}(\kbf) \\
  & = \E_q \big[ \log q(\kbf) \big] - \E_q \big[ \log p(\kbf \mid \vy) \big] \\
  & = \E_q \big[ \log q(\kbf) \big] - \E_q \big[ \log \frac{p(\vy, \kbf)}{p(\vy)} \big] \\
  & = \E_q \big[ \log q(\kbf) \big] - \E_q \big[ \log p(\vy, \kbf) \big] + \log p(\vy).
\end{aligned}
\end{equation}
The last term of the $\KLOP$ divergence, the log-marginal likelihood $\log p(\vy)$, is usually not analytically tractable. However, we use the fact that the $\KLOP$ divergence is non-negative to obtain the bound
\begin{equation}
  \log p(\vy) \geq \E_q \big[ \log p(\vy, \kbf) \big] - \E_q \big[ \log q(\kbf) \big].
\end{equation}
This inequality becomes an equality when the trial density~$q(\kbf)$  and the posterior~$p( \kbf \mid \vy) $ are equal. To minimise the $\KLOP$ divergence, it is sufficient to maximise
\mbox{$ \E_q \big[ \log p(\vy, \kbf) \big] - \E_q \big[ \log q(\kbf) \big]$,}  which is commonly referred to as the evidence lower bound (ELBO). The ELBO term can be rewritten as  
\begin{equation}
\begin{aligned}
  \ELBO(q)
  & = \E_q \big[ \log p(\vy \mid \kbf) + \log p(\kbf) \big]
    - \E_q \big[ \log q(\kbf) \big] \\
  & = \E_q\big[ \log p(\vy \mid \kbf) \big] - \kl{q(\kbf)}{p(\kbf)}.
\end{aligned}
\label{eq:elbo_equation_main}
\end{equation}
To summarise, the task now becomes:
\begin{equation} \label{eq:ELBO_max}
  q^{*}(\kbf) = \underset{q (\kbf) \in \mathcal{D}_q}{\argmax}~\E_q\big[ \log p(\vy \mid
  \kbf) \big] - \kl{q(\kbf)}{p(\kbf)}.
\end{equation}

To  maximise the $\ELBO$ with a gradient-based optimiser, we need to evaluate it and its gradients with respect to the parameters of $q(\kbf)$. 
Although the $\KLOP$ divergence term of the $\ELBO$ is often available in closed form, $\E_q\big[ \log p(\vy \mid \kbf) \big]$ involving the likelihood is generally not available. It can be approximated using a Monte Carlo approximation with $N_{\text{SVI}}$ samples from the trial density~$q(\kbf)$ as follows:
\begin{equation}
  \E_q\big[ \log p(\vy \mid \kbf) \big] \approx
  \frac{1}{N_{\text{SVI}}} \sum_{i=1}^{N_{\text{SVI}}}\log p(\vy \mid \kbf^{(i)}), \label{eq:svi_mc}
\end{equation}
where $\kbf^{(i)}$ is the $i$-th sample from $q(\kbf)$. This is done through a reparametrisation trick, as described in~\ref{sec:appendix_reparametrisationa_trick}.
Our empirical tests show that the value of $N_\text{SVI}$ in the range of 2--5 provides fast convergence of the optimisation, agreeing with previous literature~\citep{kingmaAutoencodingVariationalBayes2014}. This approach is often referred to as stochastic variational inference (SVI). 
The Monte Carlo approximation is in line with the work in~\citet{barajas-solanoApproximateBayesianModel2019} but in contrast with the analytic approximation based on the Hessian calculations proposed in~\citet{tsilifisComputationallyEfficientVariational2016}.

\subsection{Specification of trial distribution}

The specification of the approximating family of distributions determines how much structure of the true posterior distribution is captured by the variational approximation. To model complex relationships between the components of the posterior, a more complex approximating family of distributions is needed. As the richer family of distributions is likely to require more parameters, the optimisation of the usually non-convex $\ELBO$   becomes harder. A balance must be struck in this trade-off: the family should be rich enough, but the optimisation task should still be computationally tractable.

A practical and widely used variational family is the multivariate Gaussian distribution, parametrised by the mean vector and the covariance matrix. One of the key benefits of this choice is that the KL divergence term of the ELBO in \eqref{eq:elbo_equation_main} is available in closed form for a GP prior. The choice of the parametrisation of the covariance matrix determines how much structure, other than the mean estimate, is captured by the variational family. We discuss this in more detail in the next section.

Numerous approaches have been proposed to extend the trial distribution beyond the Gaussian family. A standard approach in situations when the true posterior distribution is likely to be multimodal is to consider mixtures of variational densities~\citep{bishopApproximatingPosteriorDistributions1998}. A more recent development is embedding parameters of a mean-field approximation in a hierarchical model to induce variational dependencies between latent variables \citep{tranCopulaVariationalInference2015, ranganathHierarchicalVariationalModels2016}. 

\subsubsection{Gaussian trial distribution}
\label{sec:gaussian_variationa_distribution}
%
Choosing the trial distribution  $q(\kbf)$ as a multivariate Gaussian $\mathcal{N}(\boldsymbol{\mu}, \boldsymbol\Sigma)$ requires optimisation over the mean $\boldsymbol{\mu}$ and the covariance matrix $\boldsymbol\Sigma$. The flexibility in choosing how we specify both of these parameters, especially the covariance matrix, enables us to balance the trade-off between the expressiveness of the approximating distribution and the computational efficiency. 

The richest specification corresponds to parametrising the covariance matrix $\boldsymbol{\Sigma}$ using its full Cholesky factor $\boldsymbol{L}$, i.e., 
\begin{equation}
  q(\kbf) \sim \mathcal{N}(\boldsymbol{\mu}, \boldsymbol{LL}^\top).
\end{equation} 
This choice results in a dense covariance matrix that may be able to capture the full covariance structure between the inputs (\emph{i.e.} each input may be correlated with every other input). Parametrising the components of $\boldsymbol{L}$ automatically ensures that the covariance matrix $\boldsymbol{\Sigma}$ is positive definite as necessary. The number of parameters to optimise grows as $\mathcal{O}(n_{\kappa}^2)$ and this leads to a difficult optimisation task that needs to be carefully initialised and parametrised. We refer to this parametrisation as full-covariance variational Bayes (FCVB).

A much more efficient choice is a diagonal covariance matrix, which is often referred to as mean-field variational Bayes (MFVB). By limiting the number of parameters that need to be optimised,  the optimisation task becomes simpler and the number of parameters grows only as $\mathcal{O}(n_{\kappa})$.  While more computationally efficient and easier to initialise, MFVB ignores much of the dependence structure of the posterior distribution. 

\subsection{Conditional independence and sparse precision matrices} \label{sec:method_conditional_indepdence}
%
Instead of parametrising the covariance matrix $\boldsymbol \Sigma$, or its Cholesky decomposition~$\boldsymbol L$, in physical systems it is often advantageous to parametrise the precision matrix, $\boldsymbol Q$, where $\boldsymbol{Q} = \boldsymbol{\Sigma}^{-1}$.  While a component of the covariance matrix $\mathbf{\Sigma}$ expresses \emph{marginal} dependence between the two corresponding random variables, the elements of the precision matrix reflect their \emph{conditional independence}~\citep{rueApproximateBayesianInference2009}. Or, more specifically, for two components~$\kappa_i$  and~$\kappa_j$ of a Gaussian random vector~$\kbf$ we note
\begin{equation}
	p(\kappa_i , \, \kappa_j ) = p (\kappa_i) p(\kappa_j)  \quad \Leftrightarrow \quad \Sigma_{ij}  = 0 \, , 
\end{equation}
where~$\Sigma_{ij}$ denotes the respective component of~$\boldsymbol \Sigma$. Furthermore, defining the vector~$\kbf_{-\{ i , j\}}$ from the random vector~$\kbf$ by removing its $i$-th and $j$-th component,  we note
\begin{equation}
	p(\kappa_i , \, \kappa_j  \mid  \kbf_{-\{ i , j\}} ) = p (\kappa_i \mid  \kbf_{-\{ i , j\}} ) p(\kappa_j \mid \kbf_{-\{ i , j\}})  \quad \Leftrightarrow \quad Q_{ij}  = 0 \, .
\end{equation}
That is,  ${Q}_{i j} = 0$ if and only if $\kappa_i$ is independent from $\kappa_j$, \emph{conditional} on all other components of~$\kbf$.

A succinct way to represent conditional independence is using an undirected graph whose nodes correspond to the random variables~\citep{bishopPatternRecognitionMachine2006}. A graph edge is present between two graph vertices $i$ and $j$ if the corresponding random variables are \emph{not} conditionally independent from each other, given all the other random variables. Or, expressed differently, the edges between the graph vertices correspond to non-zeros in the precision matrix.  In our context, each graph vertex represents a finite element and graph edges are introduced according to geometric adjacency of the finite elements as determined by the mesh. To this end, we define the 1-neighbourhood of a finite element as the union of the element itself and of elements sharing a node with the element. The $n$-neighbourhood is defined recursively as the union of all 1-neighbourhoods of all the elements in the $(n-1)$-neighbourhood.  We introduce an edge between two graph vertices when the respective elements are in the same $n$-neighbourhood.  

Figure~\ref{fig:methodology_conditional_independence_exmaple} shows examples of adjacency graphs and the structure of the corresponding precision matrices $\boldsymbol{Q}$ for 5 random variables resulting from a discretisation of a 1D domain with 5 finite elements. In the considered examples the random variables represent the constant log-diffusion coefficient in the elements. As shown in Figures~\ref{fig:methodology_conditional_independence_exmaple_A}  and \ref{fig:methodology_conditional_independence_exmaple_B} choosing a larger $n$-neighbourhood for graph construction leads to a denser precision matrix. For instance, from the structure of the precision matrix in Figure~\ref{fig:methodology_conditional_independence_exmaple_A}, which assumes a 1-neighbourhood structure, we can read for the log-diffusion coefficient of element~$j$ the following conditional independence relationship: 
\begin{align}
	Q_{ik} = 0   \land   Q_{il} = 0 \land Q_{im} = 0	    & \Rightarrow   p(\kappa_i \mid \kappa_j, \,  \kappa_k , \,  \, \kappa_l , \, \kappa_m   )  =  p(\kappa_i \mid  \kappa_j    )  \, .
\end{align}
When the coefficient of element~$j$ is given, the coefficient of the neighbouring element~$i$ is independent from all the remaining coefficients. This is intuitively plausible and in line with physical observations. Clearly, the covariance matrices corresponding to the given sparse precision matrices are dense. Hence, in the considered case  the coefficient of element~$i$ may still be correlated to the coefficient of element~$m$, i.e.~\mbox{ $p(\kappa_i  \mid \kappa_m) \neq p (\kappa_i)$}. This correlation will most likely be relatively weak given the large distance between the two elements, but knowing the coefficient of element~$m$ will certainly restrict the range of possible values for the coefficient of element~$i$.

\begin{figure}
    \centering
    \begin{subfigure}[b]{\textwidth}
       \centering
       \begin{minipage}{\textwidth}
       \centering
       \begin{tikzpicture}[xscale=1.0]
       \draw [thick] (0,0) -- (10,0);
       \draw (0,-.2) -- (0, .2);
       \draw (2,-.2) -- (2, .2);
       \draw (4,-.2) -- (4, .2);
       \draw (6,-.2) -- (6, .2);
       \draw (8,-.2) -- (8, .2);
       \draw (10,-.2) -- (10, .2);
       \node[align=center, above] at (1,.2){$i$};
       \node[align=center, above] at (3,.2){$j$};
       \node[align=center, above] at (5,.2){$k$};
       \node[align=center, above] at (7,.2){$l$};
       \node[align=center, above] at (9,.2){$m$};
       \end{tikzpicture}
       \end{minipage}
        \caption{Labelling of the five elements.} 
    \end{subfigure}
    \\
    \begin{subfigure}[b]{1.\textwidth}
        \centering
        \begin{minipage}{0.5\textwidth}
        \begin{tikzpicture}[auto,  node distance=1.5cm, every loop/.style={}, main node/.style={circle,draw}]
      \node[main node] (1) {$i$};
      \node[main node] (2) [right of=1] {$j$};
      \node[main node] (3) [right of=2] {$k$};
      \node[main node] (4) [right of=3] {$l$};
      \node[main node] (5) [right of=4] {$m$};
    
      \path[every node/.style={}]
        (1) edge node [] {} (2)
        (2) edge node [] {} (3)
        (3) edge node [] {} (4)
        (4) edge node [] {} (5);
    \end{tikzpicture}     
    \end{minipage}
    \begin{minipage}{0.4\textwidth}
    \centering
    $\begin{pNiceMatrix}[first-row,first-col]
          & i        & j        & k        & l      & m      \\
        i & \times   & \times   &          &        &        \\
        j & \times   & \times   & \times   &        &        \\
        k &          & \times   & \times   & \times &        \\
        l &          &          & \times   & \times & \times \\
        m &          &          &          & \times & \times     
    \end{pNiceMatrix}$
    \end{minipage}   
    \caption{Adjacency graph (left) and the corresponding adjacency matrix (right) based on 1-neighbourhood structure: there is an edge between two graph vertices if the corresponding elements share a node. \label{fig:methodology_conditional_independence_exmaple_A}}
    \end{subfigure}
    \begin{subfigure}[b]{\textwidth}
        \centering
    \begin{minipage}{0.5\textwidth}
          \begin{tikzpicture}[auto,  node distance=1.3cm, every loop/.style={}, main node/.style={circle,draw}]
      \node[main node] (1) {$i$};
      \node[main node] (2) [right of=1] {$j$};
      \node[main node] (3) [right of=2] {$k$};
      \node[main node] (4) [right of=3] {$l$};
      \node[main node] (5) [right of=4] {$m$};
    
      \path[every node/.style={}]
        (1) edge node [] {} (2)
        (1) edge[bend right=45] node [] {} (3)
        (2) edge node [] {} (3)
        (2) edge[bend right=-45] node [] {} (4)
        (3) edge node [] {} (4)
        (3) edge[bend right=45] node [] {} (5)
        (4) edge node [] {} (5);
    \end{tikzpicture}
    \end{minipage}
    \begin{minipage}{0.4\textwidth}
    \centering
    $\begin{pNiceMatrix}[first-row,first-col]
          & i        & j        & k        & l      & m      \\
        i & \times   & \times   & \times   &        &        \\
        j & \times   & \times   & \times   & \times &        \\
        k & \times   & \times   & \times   & \times & \times \\
        l &          & \times   & \times   & \times & \times \\
        m &          &          & \times   & \times & \times     
    \end{pNiceMatrix}$
    \end{minipage}
    \caption{Adjacency graph (left) and the corresponding adjacency matrix (right) based on 2-neighbourhood structure: there is an edge between two vertices if the corresponding elements are in each others 2-neighbourhoods. \label{fig:methodology_conditional_independence_exmaple_B}}
    \end{subfigure}
       \caption{An example of a 1D bar discretised with five elements and two different conditional independence assumptions.}
    \label{fig:methodology_conditional_independence_exmaple}
\end{figure}

\begin{figure}
    \centering
    \includegraphics[width=\textwidth]{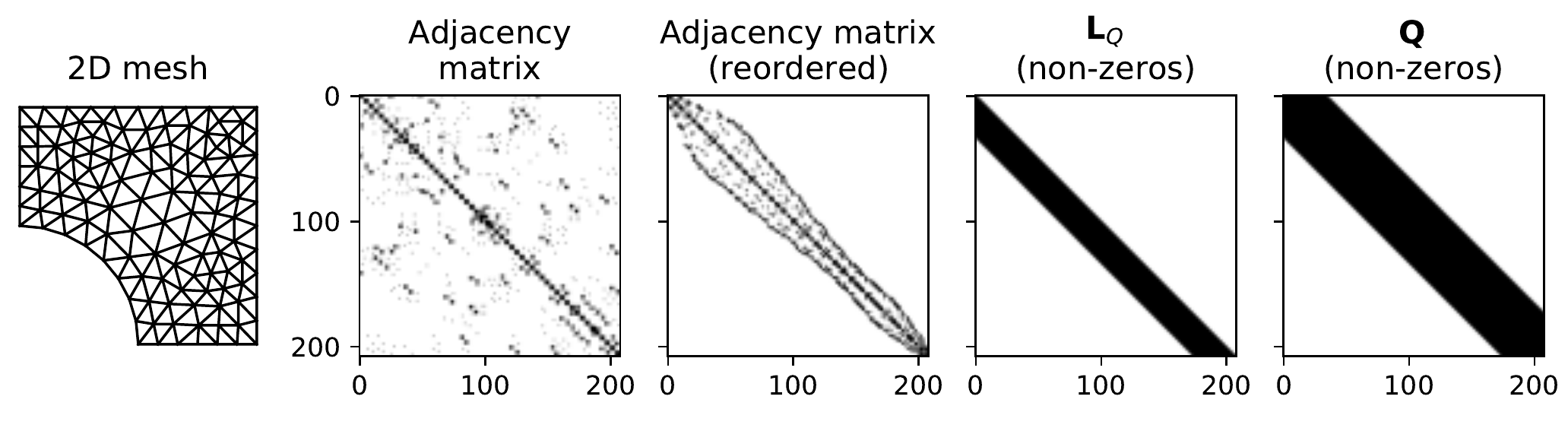}
    \caption{Sparse precision matrix parametrisation for a 2D problem. A 2-neighbourhood structure is assumed for conditional independence. The structure of the adjacency matrix depends on the specific element numbering. By renumbering the elements, one can obtain a banded adjacency matrix, which is then used to parametrise the Cholesky factor of the precision matrix, as described in Sec.~\ref{sec:gaussian_variationa_distribution}.}
    \label{fig:adjacency_precision_parametrisation}
\end{figure}

After obtaining the structure of the precision matrix, which is sparse but, in general, not banded, one can reorder the numbering of the elements in the finite element mesh to reduce its bandwidth. This allows for efficient linear algebra operations. See~\citet{cuthillReducingBandwidthSparse1969} for an example of a reordering algorithm. Once a  minimum bandwidth ordering with $b_{\text{min}}$ has been established, we use the property that the bandwidth of the Cholesky factor $\boldsymbol{L_Q}$ of matrix $\boldsymbol{Q}$ is less than or equal to the bandwidth of $\boldsymbol{Q}$~\citep{rueGaussianMarkovRandom2005}. Finally, the parameters we optimise are the components of the lower band of size $b_{\text{min}}$ of matrix $\boldsymbol{L}_Q$, so that the approximating distribution reads
\begin{equation}
  q(\kbf) \sim \mathcal{N}\Big(\boldsymbol{\mu}, (\boldsymbol{L}_Q\boldsymbol{L}_Q^\top)^{-1}\Big). \label{eq:sparse_precision_matrix_parametrisation}
\end{equation}
This process of devising a parametrisation for the precision matrix for a more complex mesh in 2D is illustrated in Fig.~\ref{fig:adjacency_precision_parametrisation}. This approach is computationally efficient -- the number of parameters grows as $\mathcal{O}(n_{\kappa})$ -- and is able to capture dependencies between all the random variables.


\subsection{Stochastic optimisation}
\label{sec:optimisation_convergence}
%
To maximise the $\ELBO$ in \eqref{eq:ELBO_max}, we use the ADAM algorithm~\citep{kingmaAdamMethodStochastic2015}. ADAM is a member of a larger class of stochastic optimisation methods that have become popular as tools for maximising non-convex cost functions. These methods construct a stochastic estimate of the gradient to perform gradient descent-based optimisation. 
ADAM, a stochastic gradient descent algorithm with an adaptive step size is one popular algorithm that exhibits a stable behaviour on many problems and is easy to use without significant tuning. The algorithm uses a per-parameter step size, which is based on the first two moments of the estimate of the gradient for each parameter. Specifically, the step size is proportional to the ratio of the exponential moving average of the 1st moment to the square root of the exponential moving average of the non-centred 2nd moment. At any point, the exponential moving average is computed with decay parameters $\beta_1$ and $\beta_2$ for the 1st and 2nd moment, respectively. We adopt the parameter values suggested in~\citet{kingmaAdamMethodStochastic2015}: $\beta_1=0.9$ and $\beta_2=0.99$. The speed of convergence is further controlled by the learning parameter $\alpha$ which is used to regulate the step size for all parameters in the same way. In our experiments, we set it to 0.01 and let it decay exponentially every 2,500 steps (1,000 for MFVB), with the decay rate of 0.96. 
While the ADAM algorithm performs well on a variety of problems, it has been shown that the convergence of this algorithm is poor on some problems~\citep{reddiConvergenceAdam2018}. We discuss alternative approaches as potential future work in Sec.~\ref{sec:conclusion}.

To monitor convergence, we use a rule that tracks an exponentially weighted moving average of the decrease in the loss values between successive steps, and stops when that average drops below a threshold. The use of such an adaptive rule gives us a way to track the convergence of the algorithm and provides a conservative estimate for the time it takes for the optimisation to converge. This rule can be adapted based on the available computational budget.
\subsection{The algorithm}
%
The maximisation of the ELBO in~\eqref{eq:elbo_equation_main} involves finding the parameters of the trial distribution $q(\kbf)$, i.e. its mean~$\boldsymbol{\mu}$ and Cholesky factor~$\boldsymbol{L_Q}$,  that minimise $\KLOP$ divergence between $q(\kbf)$ and the posterior $p(\kbf | \vy)$. Algorithm~\ref{alg:vb} shows the required steps to compute the ELBO and its gradients with respect to the parameters of the trial distribution. Different from the discussion so far, in~Algorithm~\ref{alg:vb} it is assumed that there are multiple  independent observation vectors~$\mathbf{y}_i$ with~$i \in \{ 1, \, 2, \dotsc , \, N_{\y} \}$.
%
\begin{algorithm}
	\DontPrintSemicolon 
	\KwIn{ Current parameters~$\boldsymbol{\mu}$ and~$\boldsymbol{L_Q}$ of~$q(\kbf)$}
	\KwOut{ELBO and its gradients with respect to the parameters of $q(\kbf)$}
	Sample $[\kbf^{(1)},\kbf^{(2)}, \dots, \kbf^{(N_{\text{SVI}})}]$ from $q(\kbf)$ \;
	\For{each $\kbf^{(i)}$}{
	    Solve for $\uu(\kbf^{(i)})$ and obtain gradients with respect to $\kbf$ using the FEM \;
	    $p(\y \mid \kbf^{(i)}) \gets \prod_{j=1}^{N_{\y}} p(\vy_j \mid \uu(\kbf^{(i)}), \sigma_y^2)$ and propagate its gradient with respect to $\kbf^{(i)}$ \;
	}
	ELBO $\gets N_{\text{SVI}}^{-1}\sum_{i=1}^{N_{\text{SVI}}}\log p(\vy \mid \kbf^{(i)}) + \kl{q(\kbf)}{p(\kbf)}$ and propagate the gradient with respect to the parameters of $q(\kbf)$ using the reparametrisation trick (see~\ref{sec:appendix_reparametrisationa_trick} and \citet{kingmaAutoencodingVariationalBayes2014}) \;
	\Return{\sc ELBO, $\nabla$ELBO}\;
	\caption{{ ELBO estimation and its gradient with respect to the parameters of the trial distribution.}}
	\label{alg:vb}
\end{algorithm}

\section{Examples} \label{sec:examples}

We evaluate the efficacy of variational inference first for 1D and 2D Poisson equation examples; a benchmark proposed by~\citet{aristoffBenchmarkBayesianInversion2021}; and lastly on a multimodal example of the steady-state heat equation. We discretise the examples with a standard finite element method using linear Lagrange basis functions. We compare against two sampling-based inference schemes, Hamiltonian Monte Carlo (HMC) and pre-conditioned Crank-Nicholson Markov Chain Monte Carlo (pCN); both are known to be asymptotically correct as the number of samples increases. The evaluation criteria we use focus on three aspects of an inference scheme: the accuracy with respect to capturing the mean and the variance of the solution; propagation of uncertainty in derived quantities of interest; and the time until convergence of the solution. 

To assess the propagation of uncertainty in derived quantities of interest, we consider a summary quantity for which a point estimate alone may not be informative enough for downstream tasks. In particular, we compute the log of total boundary flux through the boundary $\Gamma_{b}$:
\begin{equation}
r(\kappa)=\log \int_{\Gamma_{b}} e^{\kappa(s)} \nabla u(s) \cdot \boldsymbol{n}~\mathrm{d} s,
\end{equation}
where $\boldsymbol{n}$ is a unit vector normal to the boundary $\Gamma_{b}$. 

To quantitatively assess the inference of $\kbf$, we obtain $S$ samples from the posterior distribution of $\kbf$, $\{\kbf^{(s)}\}_{s=1}^{S}$. For synthetic experiments, where we know the true $\kbf$ which generated the observations, we compute the mean $\kbf$ error norm. The computation is the Euclidean norm of the error between the true value, $\kbf_\text{true}$, and the mean of the obtained samples: 
\begin{equation} \label{eq:computations_mean_kappa_error}
    \text{Mean } \kbf \text{ error} = \bigg\Vert \frac{1}{S}\sum_{s=1}^S  \kbf^{(s)} - \kbf_\text{true} \bigg\Vert_{2}.
\end{equation}
Further, we compute the expected error in the solution space. This measures how close the solutions corresponding to the samples of $\kbf$ are to the true solution $\uu(\kbf_\text{true})$. Specifically, we compute
\begin{equation} \label{eq:computations_expected_solution_error}
    \text{Mean } \uu(\kbf) \text{ error} = \frac{1}{S}\sum_{s=1}^S \big\Vert \uu(\kbf^{(s)}) - \uu(\kbf_\text{true}) \big\Vert_2.
\end{equation}

\subsection{Poisson 1D}\label{sec:1d-poisson}

For this experiment, we assume the unit-line domain, which is discretised into 32 equal-length elements. We impose Dirichlet boundary conditions on both boundaries, specifically we set $u(0)=u(1)=0$; the forcing is constant everywhere $f(\x)=1$. 
Unless specified otherwise, all experiments in this section use $N_{\y}=5$ observations per sensor and the sensor noise $\sigma_y = 0.01$. Sensors are located on each of the discretisation nodes. For the prior on $\kappa$, we choose a zero-mean Gaussian process with squared exponential kernel (see Sec.~\ref{sec:background-prior-discussion} for details). We compare the results for three specifications of the prior length-scale, \mbox{$\ell_\kappa \in \{ 0.1, 0.2, 0.3\}$}. The length-scale used to generate the data is $\ell_\kappa = 0.2$. For inferences made using data generated by a shorter length-scale, see~\ref{sec:short_lengthscale_results}.

\subsubsection{VB performs competitively based on error norms}

\begin{figure}
\centering
\includegraphics[width=.485\textwidth]{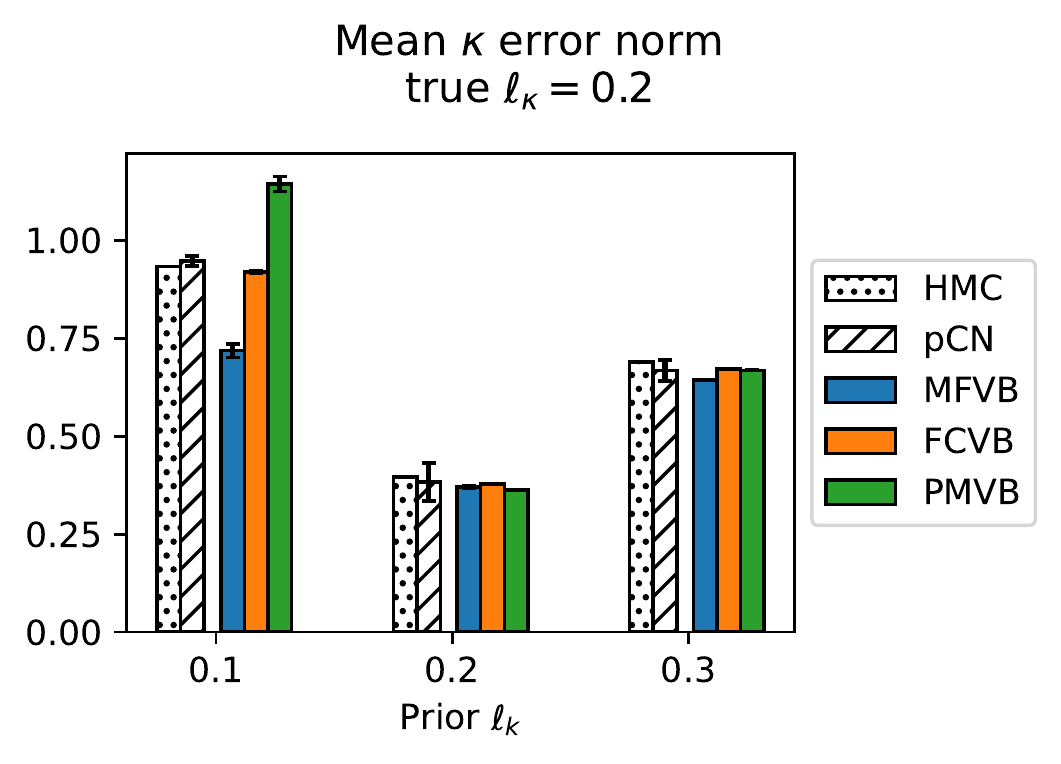} 
\includegraphics[width=.485\textwidth]{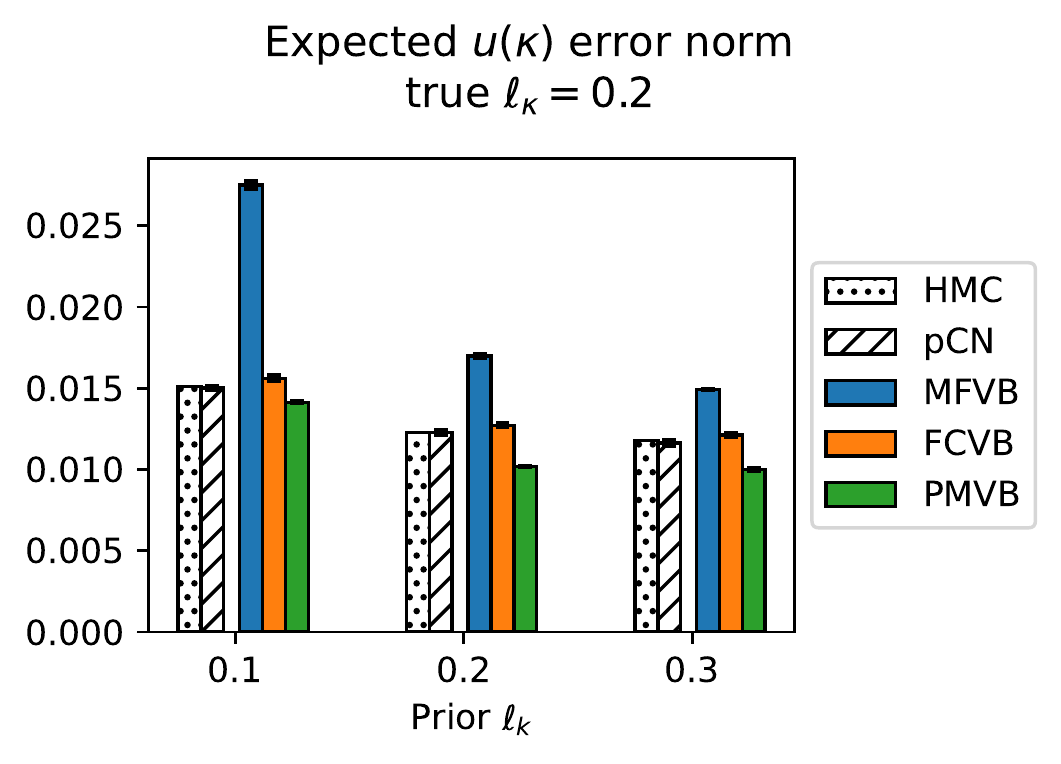} 
\caption{Mean $\kbf$ error norm for the Poisson 1D problem (left), as defined in~\eqref{eq:computations_mean_kappa_error}, and expected solution error norm (right), as defined in~\eqref{eq:computations_expected_solution_error}. Both quantities are estimated using 10,000 samples from the inferred posterior distribution of $\kbf$. Quantitatively, the sampling methods (HMC and pCN) and VB produce comparable results in both metrics, except MFVB parametrisation which captures the mean of $\kbf$ very well, but fails to account for the uncertainty as manifested in high error norm in the solution space. For a qualitative comparison, see Fig.~\ref{fig:poisson1d_main} where each row of results corresponds to a different value of the true prior length-scale $\ell_\kappa$.}
\label{fig:poisson1d_score_and_error}
\end{figure}

Fig.~\ref{fig:poisson1d_score_and_error} shows the mean $\kbf$ error norm~\eqref{eq:computations_mean_kappa_error} and the expected solution error norm~\eqref{eq:computations_expected_solution_error} obtained from 10,000 posterior samples of $\kbf$ from Hamiltonian Monte Carlo (HMC), pre-conditioned Crank-Nicholson MCMC (pCN), as well as VB inference with different parametrisations of the covariance/precision matrix.
It is evident that for prior length-scales \mbox{$\ell_\kappa \in \{0.2, 0.3\}$}, the mean $\kbf$ error norms computed by the variational Bayes methods are very close to the estimates from HMC and pCN. For prior $\ell_\kappa = 0.1$, the mean $\kbf$ error norm computed by MFVB is lower than other VB methods and MCMC methods. This is most likely due to MFVB being a much easier optimisation task compared to other VB methods with more optimisation parameters that capture dependencies.  For the expected solution error norm, MFVB posterior consistently underestimates the uncertainty in $\kbf$, thus ignoring possible values of $\kbf$ which are consistent with the data. This is further confirmed in the qualitative assessment of uncertainty in the next section. While MCMC methods are asymptotically correct, in practice, devising efficient samplers for high-dimensional problems within a limited computational budget is still a challenging task and requires substantial hand-tuning. To affirm that all the VB methods provide a good estimate of the mean of $\kbf$, as compared to MCMC methods, is better demonstrated by inspecting Fig.~\ref{fig:poisson1d_main} which shows not only the mean but also the posterior uncertainty, which we discuss next.

\subsubsection{VB adequately estimates posterior variance}
\begin{figure}
\centering
\includegraphics[width=0.8\textwidth]{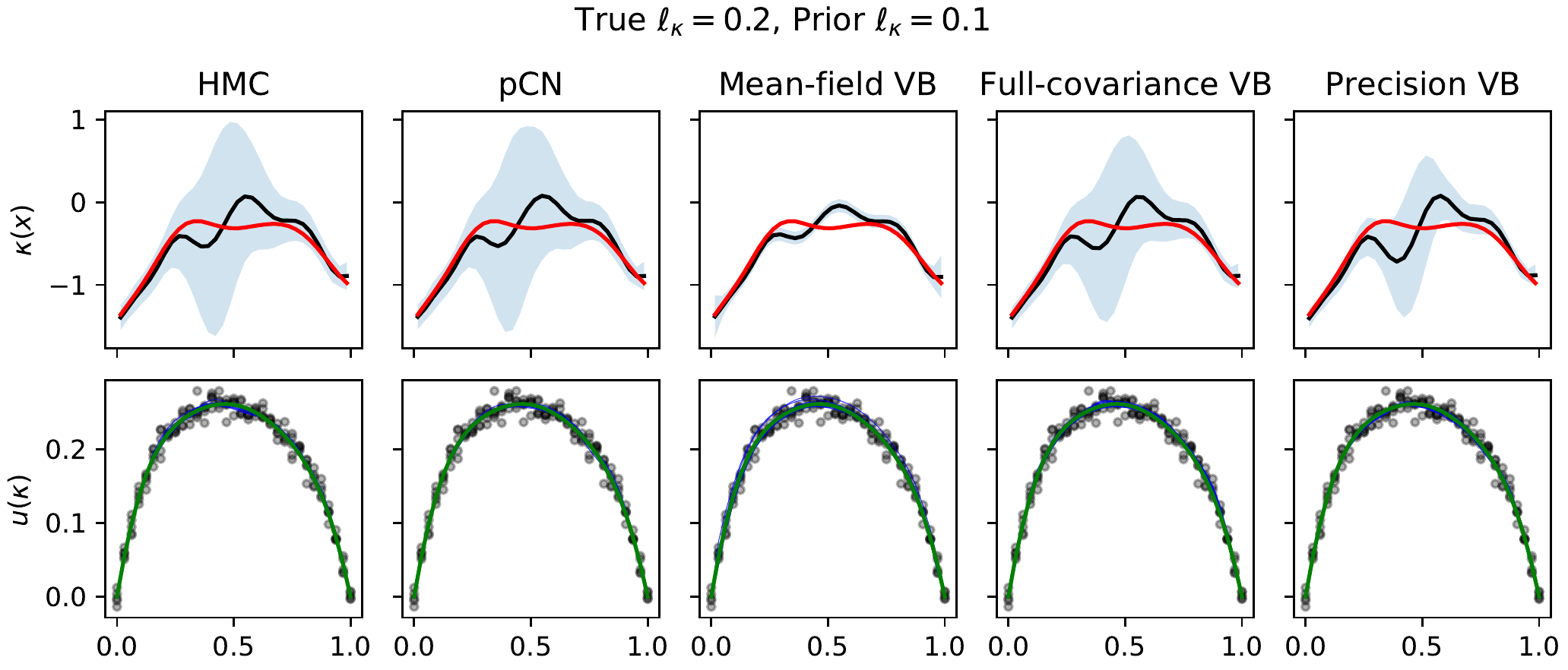}
\\
\vspace{1.2em}
\includegraphics[width=0.8\textwidth]{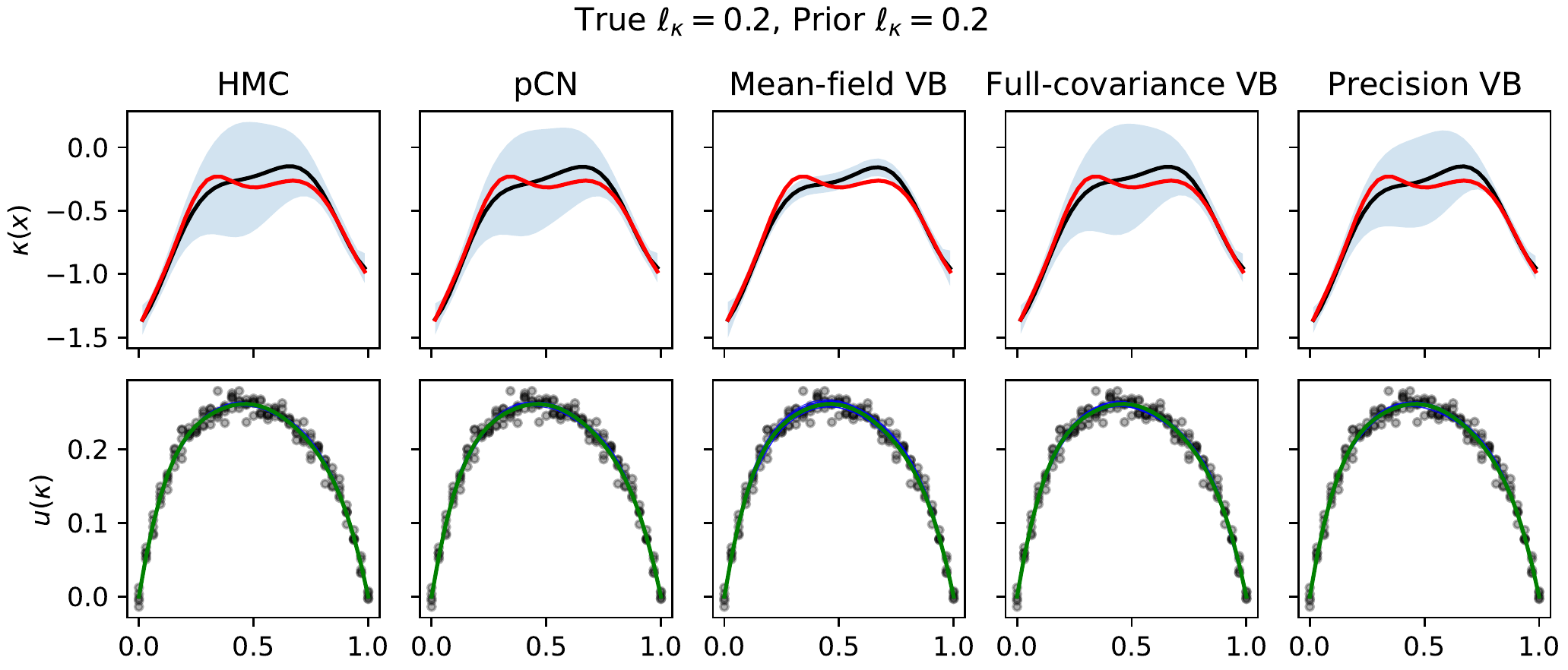}
\\
\vspace{1.2em}
\includegraphics[width=0.8\textwidth]{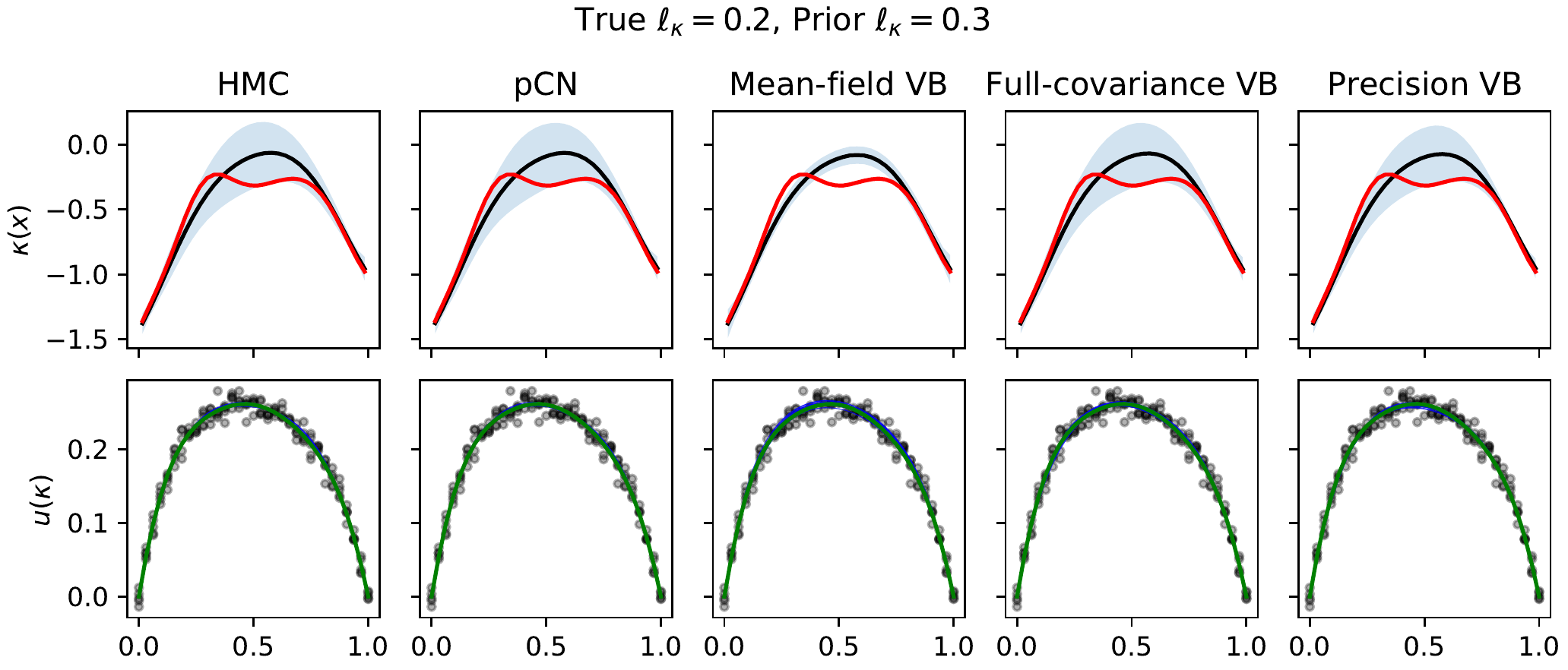}
\\
\includegraphics[width=0.8\textwidth]{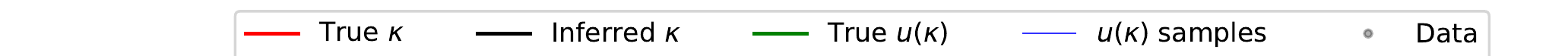}
\caption{Top row in each of the three panels show true values of $\kbf$ (red), posterior means (black) and plus and minus two times the standard deviation (blue shaded regions) for HMC, pCN, and VB variants for different values of prior length-scales $\ell_{\kappa}$. The bottom rows show the data (black), true solution $\uu$ (green), solutions for different samples of $\kappa$ (blue). For the PMVB estimate, the bandwidth is set to $10$.}
\label{fig:poisson1d_main}
\end{figure}

Figure~\ref{fig:poisson1d_main} shows the true values of $\kbf$ (red), the posterior means (black) and plus and minus two times the standard deviation (blue shaded regions) estimated by HMC, pCN, and variational inference with mean-field (MFVB), full covariance (FCVB), and precision matrix (PMVB) parametrisations for different values of prior length-scales.
\begin{figure}
    \centering
    \includegraphics[width=\textwidth]{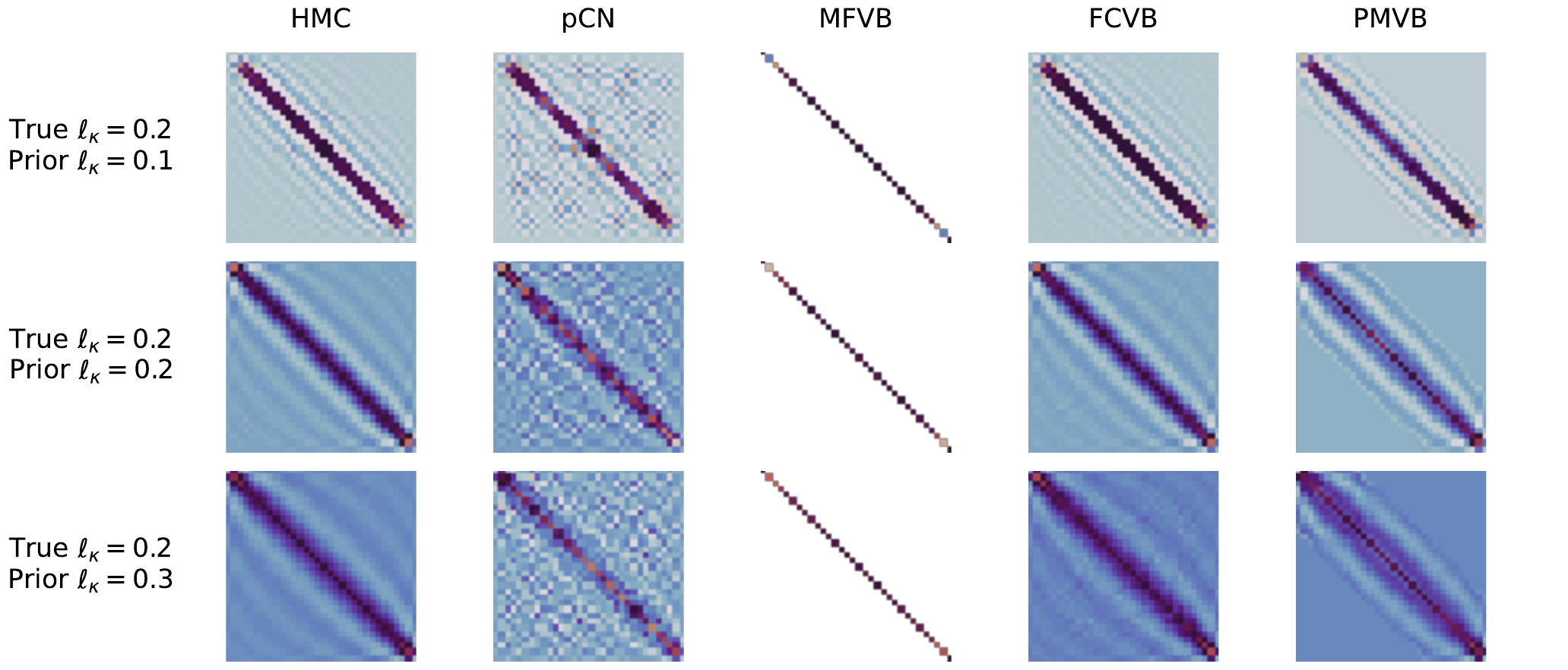}
    \caption{Precision matrices for each of the considered methods, where true $\ell_\kappa = 0.2$ and each row corresponds to a different value of prior $\ell_\kappa$.}
    \label{fig:poisson1d-precision-matrix-sparsity-structure}
\end{figure}
We observe that the posterior variance estimates computed by HMC, pCN, and full covariance VB are qualitatively very similar, with the estimated uncertainty increasing with increasing distance from the fixed boundary. However, the MFVB solution greatly underestimates posterior variance while computing a reasonable estimate of the posterior mean. The over-confidence of MFVB means that values of $\kbf$ that are consistent with the observed data are ignored; this may lead to poor calibration if the MFVB posterior is used as the true $\kbf$ in downstream tasks or in other contexts. For the PMVB parametrisation, the uncertainty is underestimated to a much lesser extent. 

To demonstrate the dependence structure captured by each method, Fig.~\ref{fig:poisson1d-precision-matrix-sparsity-structure} shows the heatmap of the corresponding precision matrices. Visual inspection suggests that the precision structure inferred using FCVB matches that of MCMC methods while MFVB does not consider covariance relationships by design. The PMVB parametrisation, that takes into account the structure of the problem offers a trade-off between capturing the majority of the correlations in the problem but allowing for more efficient inference due to the sparsity in the resulting matrix. Qualitatively, the PMVB uses only a fraction of the entries in the precision matrix in comparison to the FCVB while consistently achieving a similar ELBO, as demonstrated in Fig.~\ref{fig:loss_traceplot_1d_num_svi_samples}.

\begin{figure}
    \centering
    \includegraphics[width=0.8\textwidth]{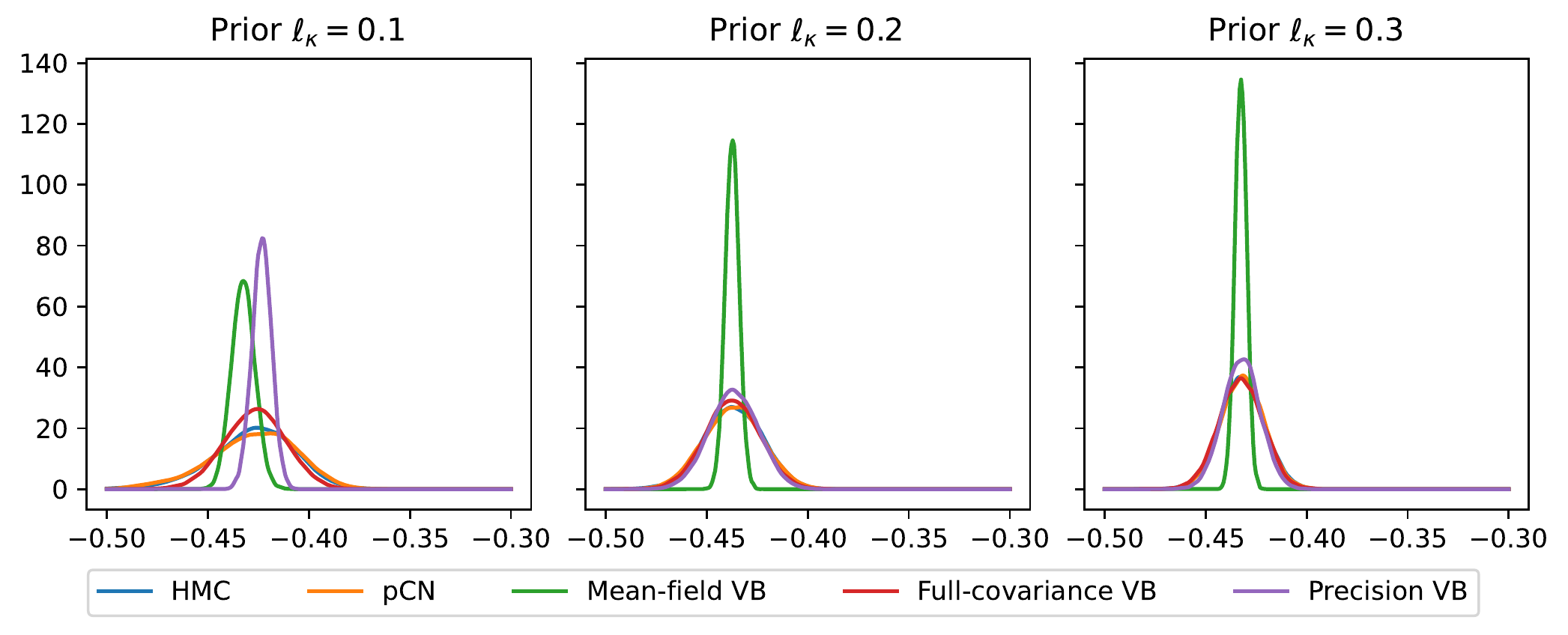}
    \caption{Log of the boundary flux at the left boundary node (\mbox{$x=0$}) for the 1D Poisson example. For PMVB, the precision matrix bandwidth of 10 is used.}
    \label{fig:results-1d-qoi-total-flux}
\end{figure}

The observations above are further confirmed by the density plot of our quantity of interest: the log of the total flux on the boundary, shown in Fig.~\ref{fig:results-1d-qoi-total-flux}. 
For this example, we compute the flux on the left boundary at $x=0$ and show the posterior distribution of this quantity. For longer prior length-scales, FCVB and PMVB agree with the estimates obtained from pCN and HMC, whereas mean-field VB underestimates the uncertainty. For the short prior length-scale ($\ell_{\kappa}=0.1$), both PMVB and MFVB underestimate the uncertainty as compared with HMC, pCN, and FCVB schemes. The posterior distribution of FCVB approximately agrees with the MCMC schemes.

\begin{figure}
    \centering
    \includegraphics[width=0.8\textwidth]{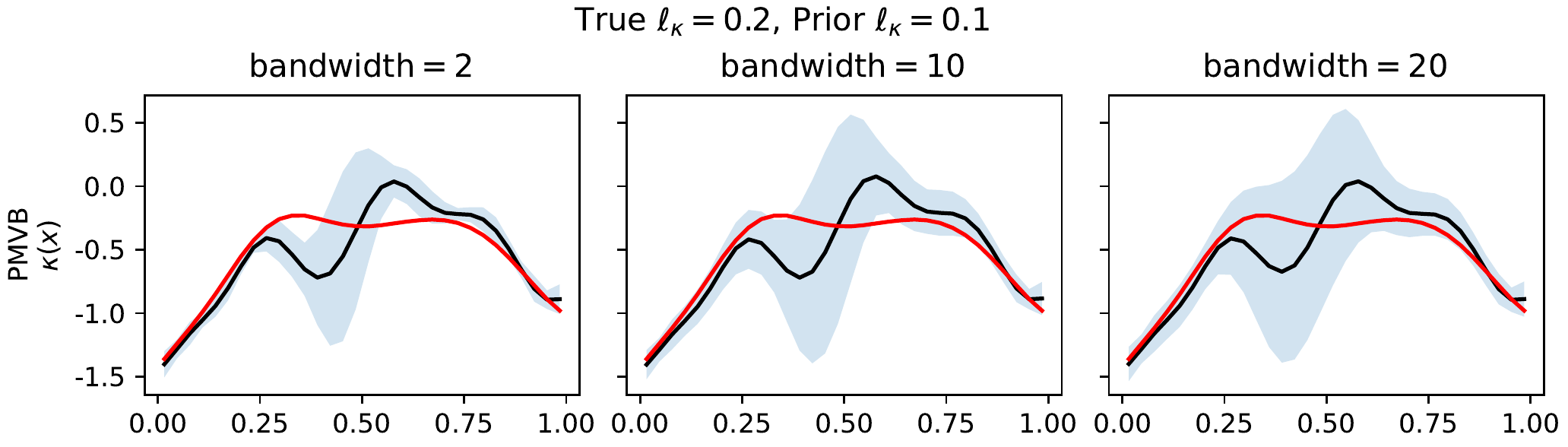}
    \\
    \vspace{1.2em}
    \includegraphics[width=0.8\textwidth]{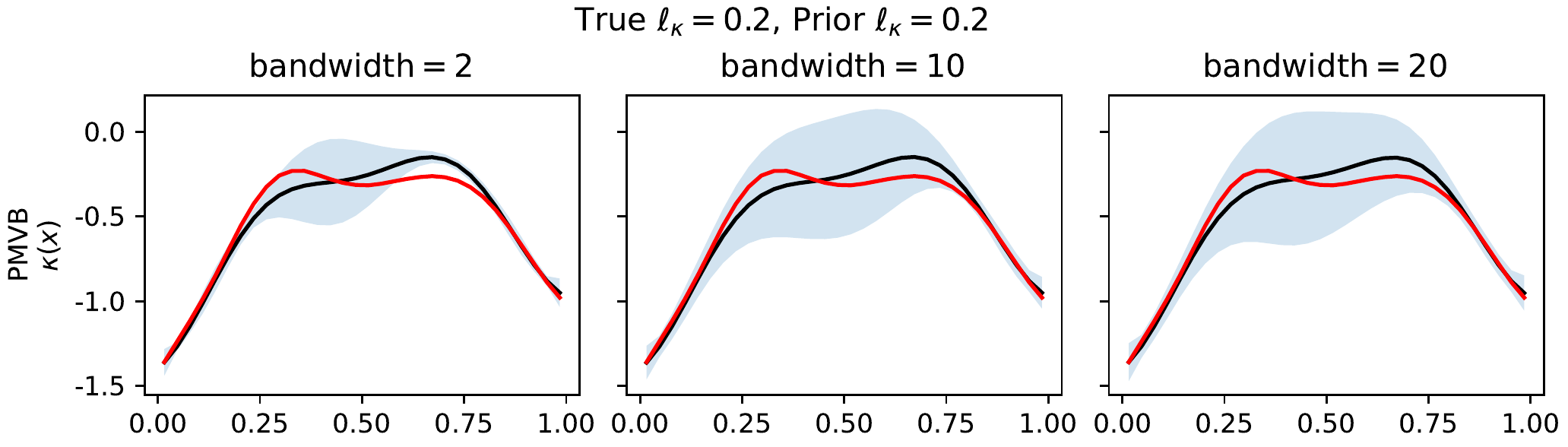}
    \\
    \vspace{1.2em}
    \includegraphics[width=0.8\textwidth]{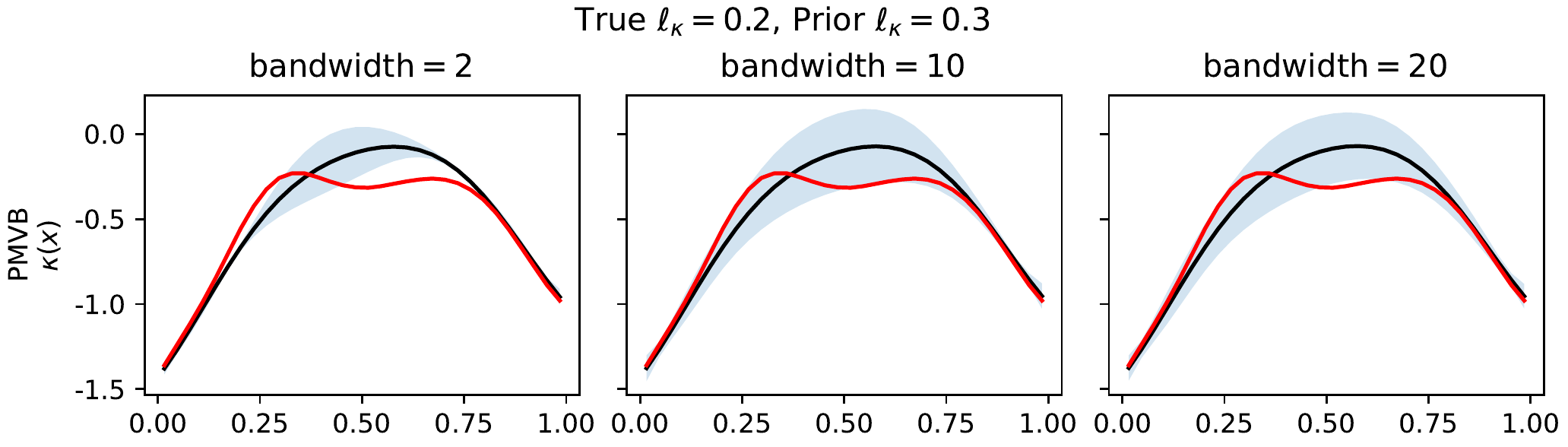}
    \caption{True values of $\kbf$~(red), posterior means~(black) and plus and minus two times the standard deviation~(blue shaded region) for different matrix bandwidths of the precision matrix parametrisation of VB. Bandwidth corresponds to the order of neighbourhood structure considered when parametrising $\boldsymbol{Q}.$}
    \label{fig:results-1d-pmvb-multi-bandiwdth}
\end{figure}

For the results obtained using the PMVB scheme, we used the 10-neighborhood structure to define the adjacency matrix and the non-zero elements of the precision matrix, $\boldsymbol{Q}$ (see Sec.~\ref{sec:method_conditional_indepdence}). The order of the neighbourhood structure, which corresponds to the precision matrix bandwidth, determines how much dependence within $\kbf$ is captured by the approximating posterior distribution. In Fig.~\ref{fig:results-1d-pmvb-multi-bandiwdth}, we show how the estimate of the mean and the variance of $\kbf$ changes for different orders of neighbourhood structure. As expected, with the increasing bandwidth, the posterior estimate of $\kbf$ gets closer to the estimate of FCVB, HMC, and pCN (shown in Fig.~\ref{fig:poisson1d_main}). While there is a significant change in the uncertainty estimate when we increase the bandwidth from 2 to 10, it is less pronounced when we change it from 10 to 20. For this reason, we choose the value of 10 for the PMVB parametrisation in 1D.

\subsubsection{VB estimates improve with more observations and decreasing observational noise}

The consistency of the posterior refers to the contraction of the posterior distribution to the truth as the data quality increases, \emph{i.e.} either the number of observations increases or observation noise tends to zero. A recent line of work~\citep{abrahamStatisticalCalderonProblems2020, monardStatisticalGuaranteesBayesian2020,giordanoConsistencyBayesianInference2020} showed the posterior consistency for the estimates obtained using popular MCMC schemes such as pCN or unadjusted discretised Langevin algorithm for Bayesian inverse problems based on PDE forward mappings. While similar results are not available for VB methods in infinite-dimensional case, consistency and Bernstein-von Mises type results have been shown for the finite-dimensional case, including Bayesian inverse problems~\citep{wangFrequentistConsistencyVariational2019, luGaussianApproximationsProbability2017}. Empirically, our experiments show that for the given family of trial distributions the VB posterior distribution contracts to the true $\kbf$.

\begin{figure}
    \centering
    \includegraphics[width=0.7\textwidth]{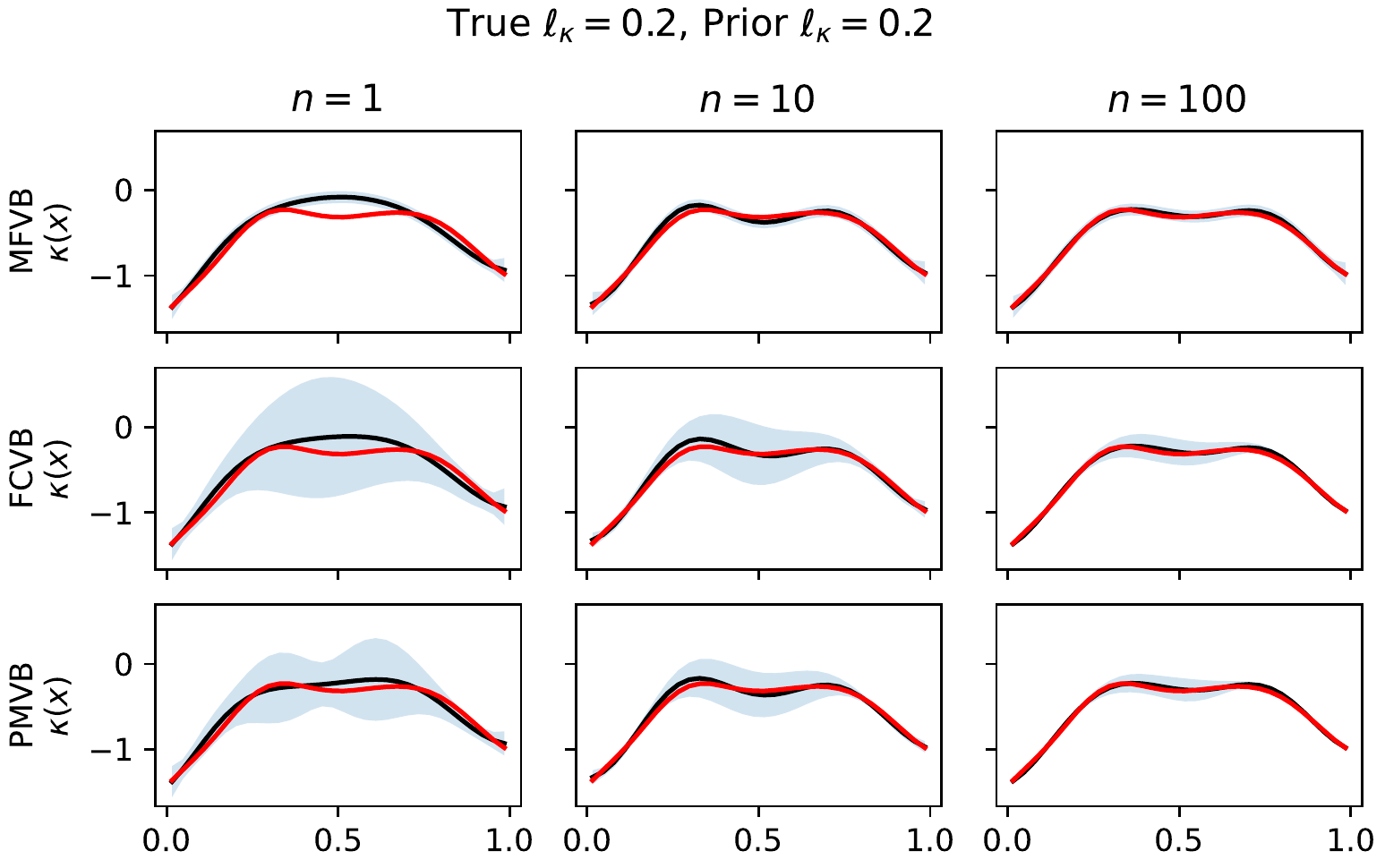}
    \\
    \vspace{1.2em}
    \includegraphics[width=0.7\textwidth]{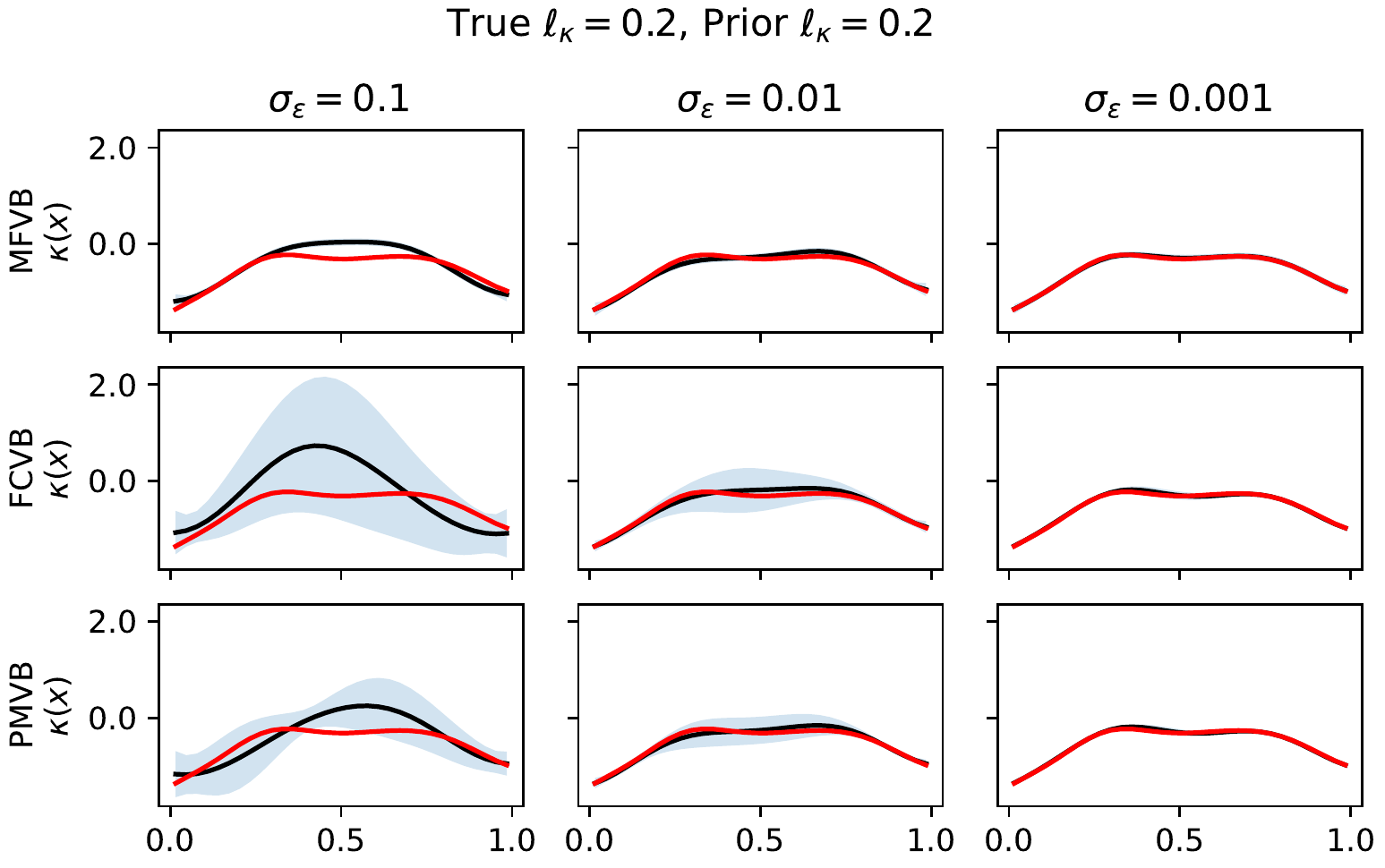}
    \caption{True values of $\kbf$ (red), posterior means (black) and plus and minus two times the standard deviation (blue shaded regions) for VB with different parametrisations for different number of observations per sensor, $N_{\y} \in \{ 1, 10, 100\}$ (top panel), and for different values of sensor noise $\sigma_{\epsilon} \in \{ 0.1, 0.01, 0.001\}$ (bottom panel).}
    \label{fig:poisson1d_uncertainty_num_points_noise}
\end{figure}

Firstly, we show that increasing the number of observations, $N_{\y}$, results in a more accurate estimate. Given that the observations, $\{ \vy_i\}_{i=1}^{N_{\y}}$, are independent of each other, the likelihood term of the ELBO (see Eq. \eqref{eq:elbo_equation_main}) is the product of the individual likelihood terms:
\begin{equation}
p(\vy_1, \dots, \vy_{N_{\y}} \mid \kbf) = \prod_i^{N_{\y}} p(\vy_i \mid \kbf).
\end{equation}
Secondly, by decreasing the observational noise $\sigma_\y$ we expect the posterior distribution to get closer to the ground truth and with lower uncertainty. Fig.~\ref{fig:poisson1d_uncertainty_num_points_noise} shows the true values of $\kbf$ (red), the posterior mean estimates (black) and plus and minus two times the standard deviation (blue shaded regions) obtained by different variants of variational Bayes for varying numbers of observations (top panel) and different values of observational noise (bottom panel). We can see that MFVB underestimates the posterior variances and these estimates do not depend on the number of observations (top panel in Fig.~\ref{fig:poisson1d_uncertainty_num_points_noise}) or the amount of observational noise (bottom panel in Fig.~\ref{fig:poisson1d_uncertainty_num_points_noise}). However, the FCVB and PMVB uncertainty estimates get narrower with increasing number of observations and with decreasing observational noise, which is a desirable behaviour that should be exhibited by any consistent uncertainty estimation method. We can also see that the true solution is contained within the uncertainty bounds for all numbers of observations and noise levels for the full covariance parametrisation. This is not the case for the mean-field VB, providing another indication of uncertainty underestimation for this parametrisation.

\subsubsection{VB is an order of magnitude faster than HMC}

\begin{figure}
    \centering
    \includegraphics[width=0.8\textwidth]{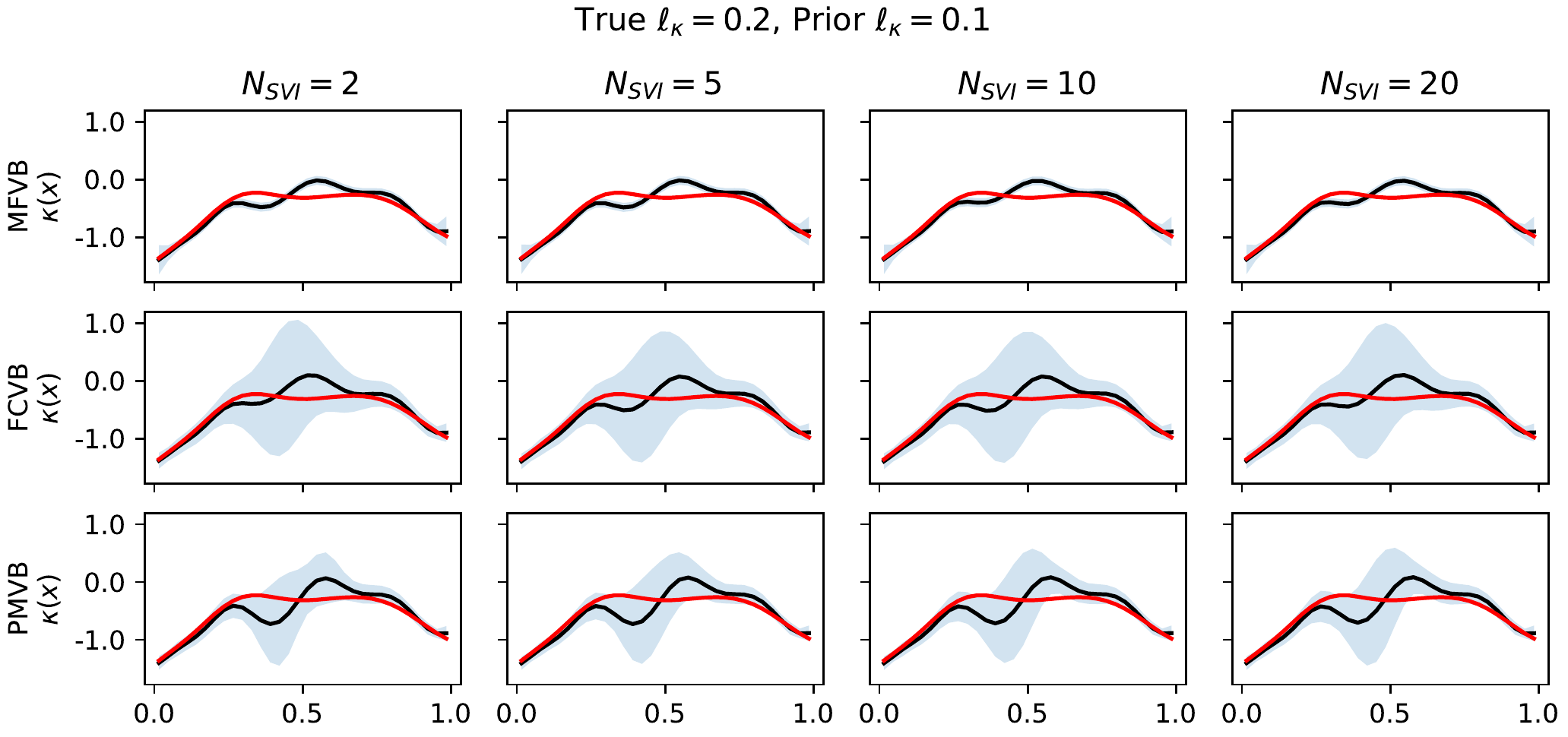}
    \\
    \vspace{1.2em}
    \includegraphics[width=0.8\textwidth]{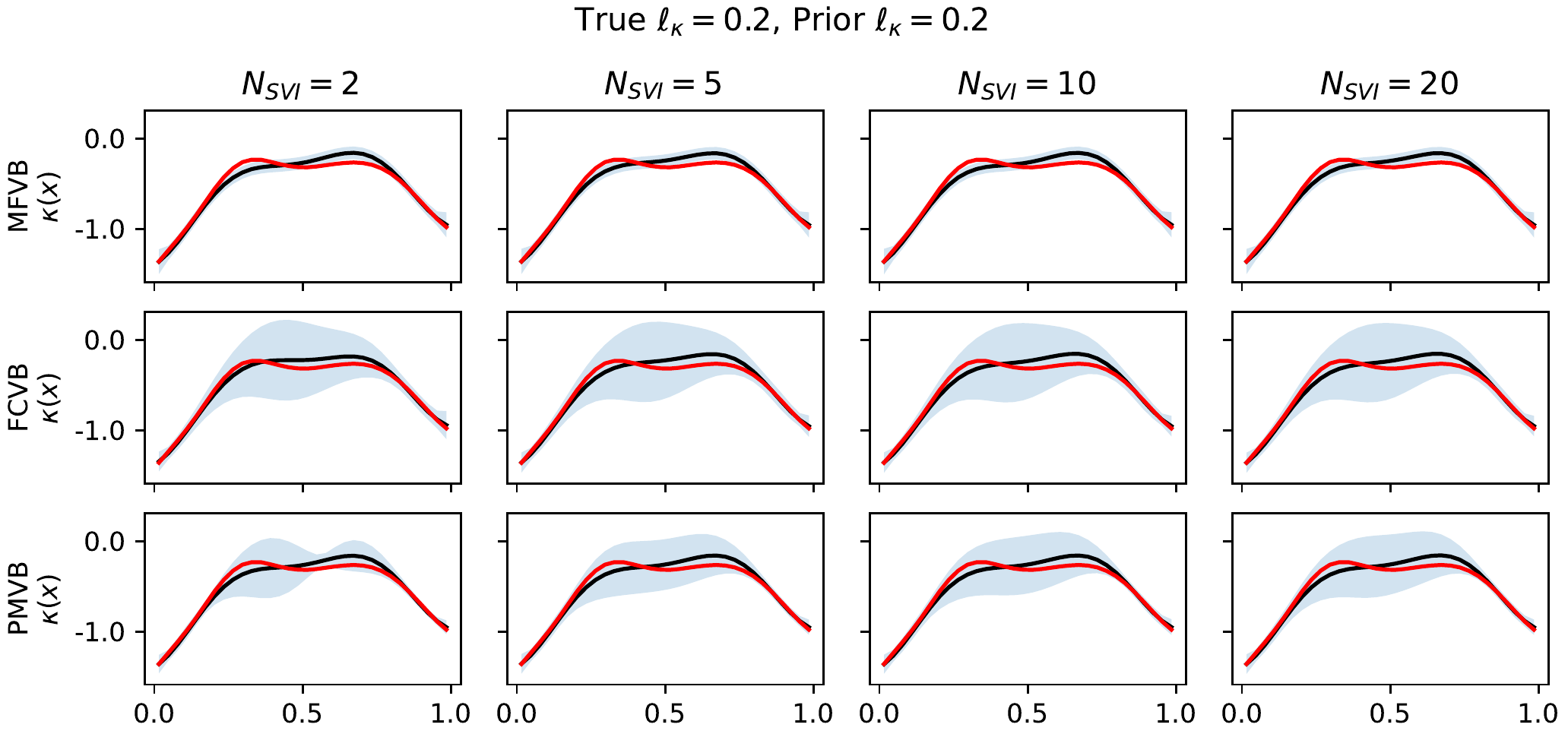}
    \\
    \vspace{1.2em}
    \includegraphics[width=0.8\textwidth]{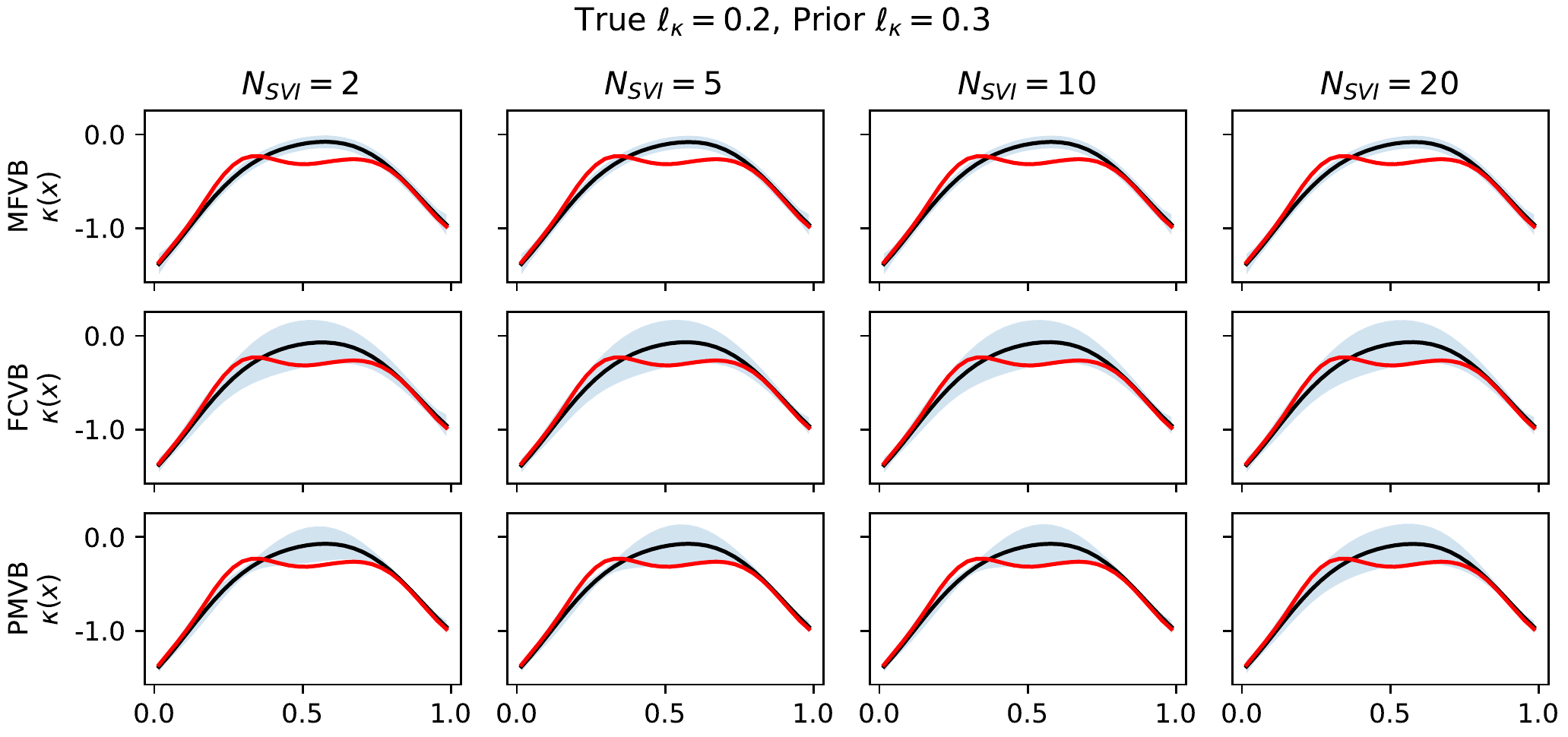}
    \caption{True values of $\kbf$ (red), posterior means (black)  and plus and minus two times the standard deviation (blue shaded regions) of VB with different parametrisations for varying number of Monte Carlo samples when computing ELBO. Three different length-scales for the prior are shown: 0.1, 0.2, 0.3.}
    \label{fig:poisson1d_uncertainty_num_svi_samples}
\end{figure}

\begin{figure}
    \centering
    \includegraphics[width=1.0\textwidth]{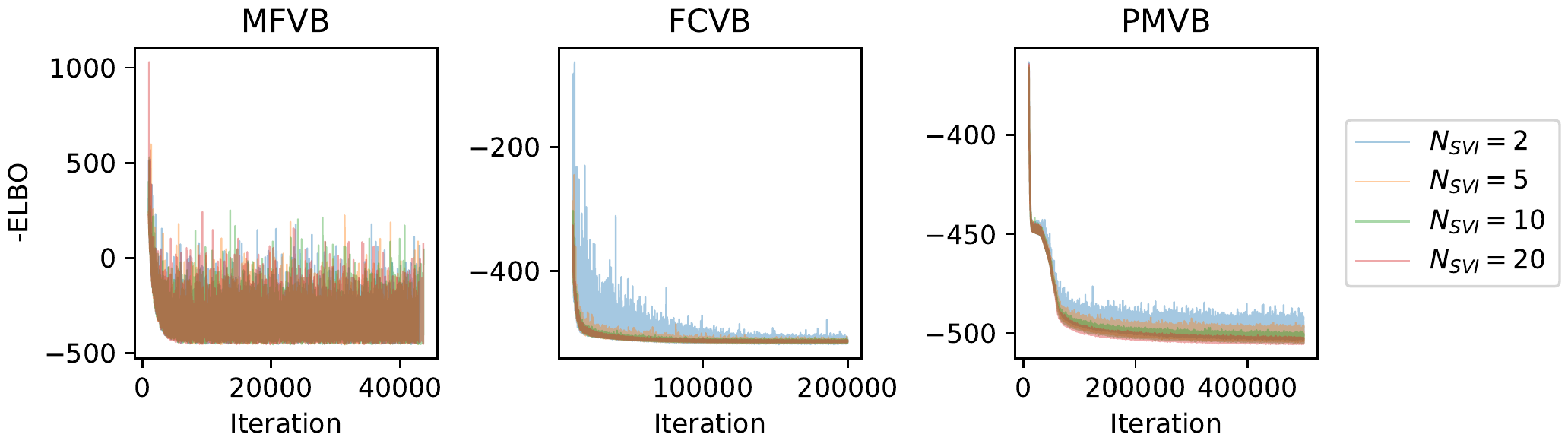}
    \caption{Negative ELBO trace plot for both MFVB and FCVB for different values of $N_\text{SVI}$. For this example, true $\ell_\kappa=0.2$ and prior $\ell_\kappa=0.1$.}
    \label{fig:loss_traceplot_1d_num_svi_samples}
\end{figure}

For HMC estimates, we obtain 200,000 samples out of which the first 100,000 are used to calibrate the sampling scheme and are subsequently discarded. Table~\ref{tab:poisson_1d_run_times} provides the run-times for HMC, MFVB, FCVB, and PMVB. For the HMC column, we also report (shown in brackets) the range of effective sample sizes (ESS) across different components of $\kbf$. For details on ESS, we refer the reader to~\cite[Ch. 11]{gelmanBayesianDataAnalysis2013}. Even with conservative convergence criteria (described in Sec.~\ref{sec:optimisation_convergence}), the computational cost of VB algorithms is up to 25 times lower than that of HMC. 
\begin{table}
    \centering
    \begin{tabular}{llrrrrr}
    \toprule
    \multirow{2}{*}{true $\ell_\kappa$} & \multirow{2}{*}{prior $\ell_\kappa$} & \multicolumn{5}{c}{Time (hours)} \\
     \cmidrule{3-7}
        &     & \multicolumn{2}{c}{HMC}            &         MFVB &         FCVB   & PMVB \\
    \cmidrule{1-2} \cmidrule{3-7}
0.1 & 0.1 & 15.2 & (871--3244)  &  1.1 &  3.6 &  2.1 \\
    & 0.2 & 11.1 & (1043--4006) &  0.7 &  2.7 &  2.1 \\
    & 0.3 &  7.2 & (1130--5408) &  0.6 &  2.3 &  2.0 \\
0.2 & 0.1 & 15.2 & (1600--4700) &  0.6 &  2.2 &  1.8 \\
    & 0.2 & 10.4 & (1067--3468) &  0.6 &  2.3 &  2.0 \\
    & 0.3 &  7.0 & (1487--3969) &  0.5 &  1.7 &  1.8 \\
\bottomrule 
    \end{tabular}
    \caption{Run-times for different inference schemes in hours for the Poisson 1D problem. For VB methods, $N_\text{SVI}=3$. The column for HMC includes the range of effective sample sizes (ESS) across different components of $\kbf$.}
    \label{tab:poisson_1d_run_times}
\end{table}
To emphasise the computational efficiency of the variational inference, we show the posterior estimates for different number of Monte Carlo samples in the estimation of ELBO. Fig.~\ref{fig:poisson1d_uncertainty_num_svi_samples} shows that on a qualitative level, a low number of samples is sufficient to obtain a good estimate.
In particular, even with 2 Monte Carlo samples, the estimates are very similar to the case where $N_\text{SVI}=20$. However, a lower number of samples may result in slower convergence of the  optimisation scheme. Fig.~\ref{fig:loss_traceplot_1d_num_svi_samples} shows that for the FCVB and PMVB parametrisations, where the number of optimised parameters is larger than for MFVB, increasing the number of SVI samples may speed up the convergence of the optimisation. The effect is not as strong for the MFVB parametrisation.

\subsection{Poisson 2D} 

\begin{figure}
\centering
\begin{minipage}{.45\textwidth}
    \centering
    \includegraphics[height=0.65\textwidth]{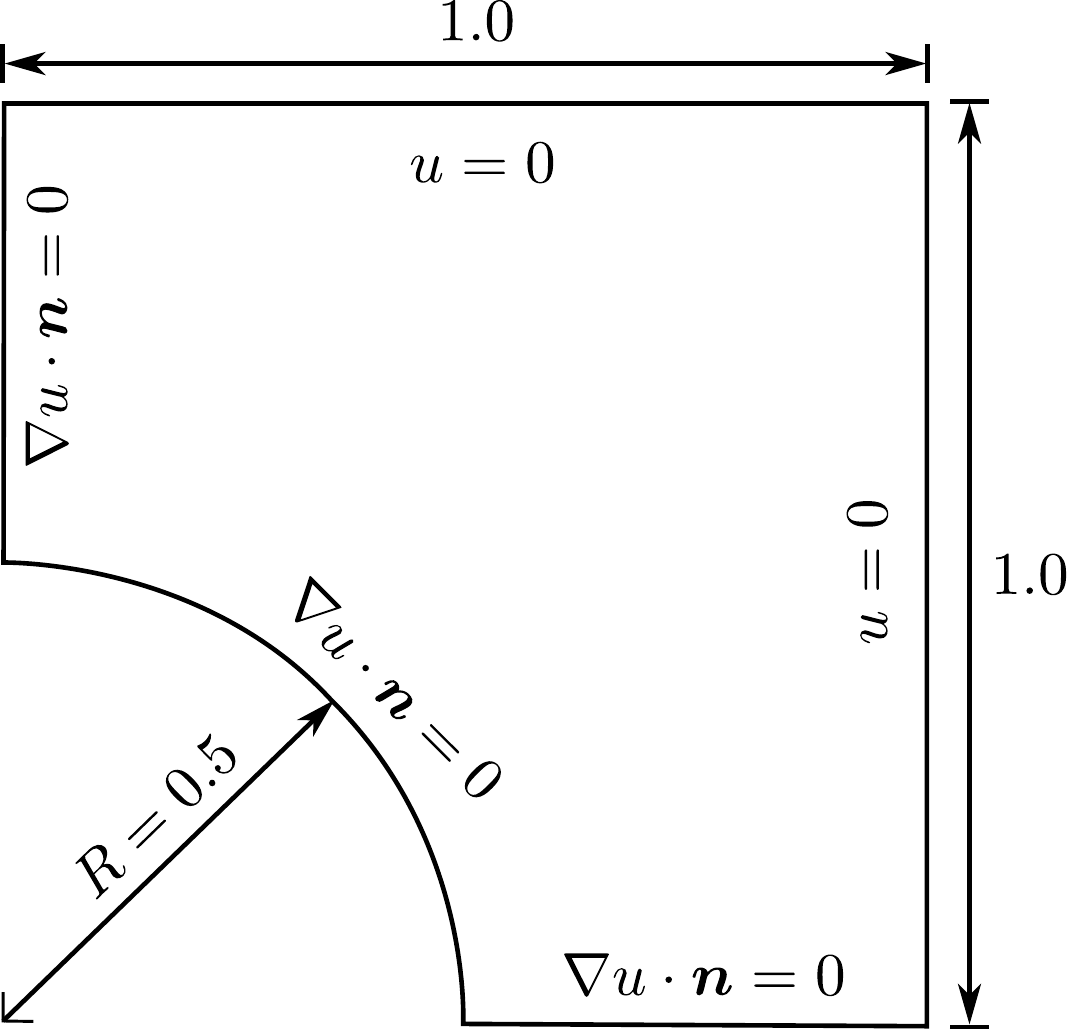}
\end{minipage}%
\begin{minipage}{0.45\textwidth}
    \centering
    \includegraphics[height=0.78\textwidth]{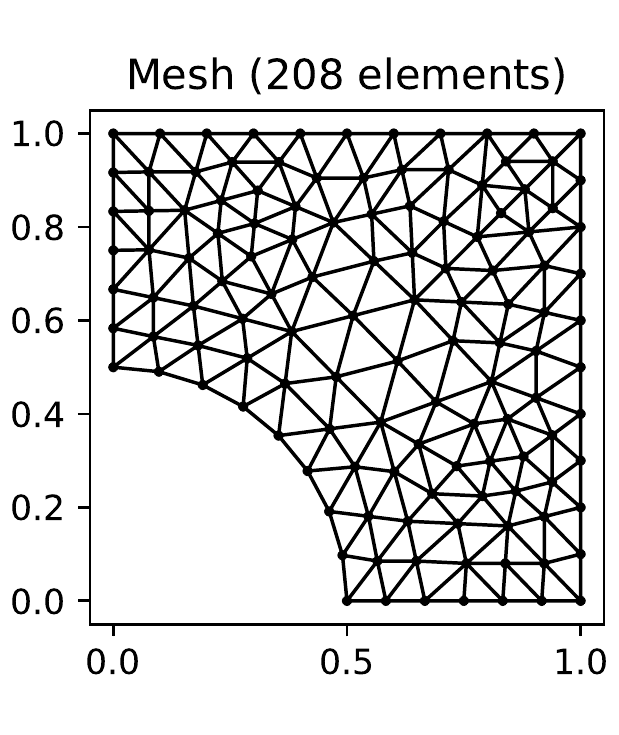}
\end{minipage}
\caption{Left: Specification of the domain for the 2D Poisson problem. Note that we impose Dirichlet boundary conditions $u(x, y)=0$ when $x=1$ or $y=1$. We impose Neumann boundary conditions on the rest of the boundary. Right: a triangular discretisation of the domain.}
\label{fig:poisson2d_domain}
\end{figure}

We consider a 2D Poisson problem on the unit-square domain with a circular hole as shown in Fig.~\ref{fig:poisson2d_domain}, with boundary conditions as indicated in the same figure. The problem is discretised with 208 linear triangular elements and 125 nodes. The forcing term is assumed to be constant throughout the domain, $f(\vx)=1$. Unless specified otherwise, all experiments in this section use $N_{\y}=5$ observations per sensor and the sensor noise $\sigma_y = 0.001$ (note that for the 1D example we used $\sigma_y = 0.01$). The sensors are located at each node of the mesh. As in the 1D example, we assume a zero-mean GP prior on $\kappa$ with square exponential kernel with varying length-scale, $\ell_\kappa$, as discussed in Sec.~\ref{sec:background-prior-discussion}. 

\begin{figure}
    \centering
    \includegraphics[width=0.485\textwidth]{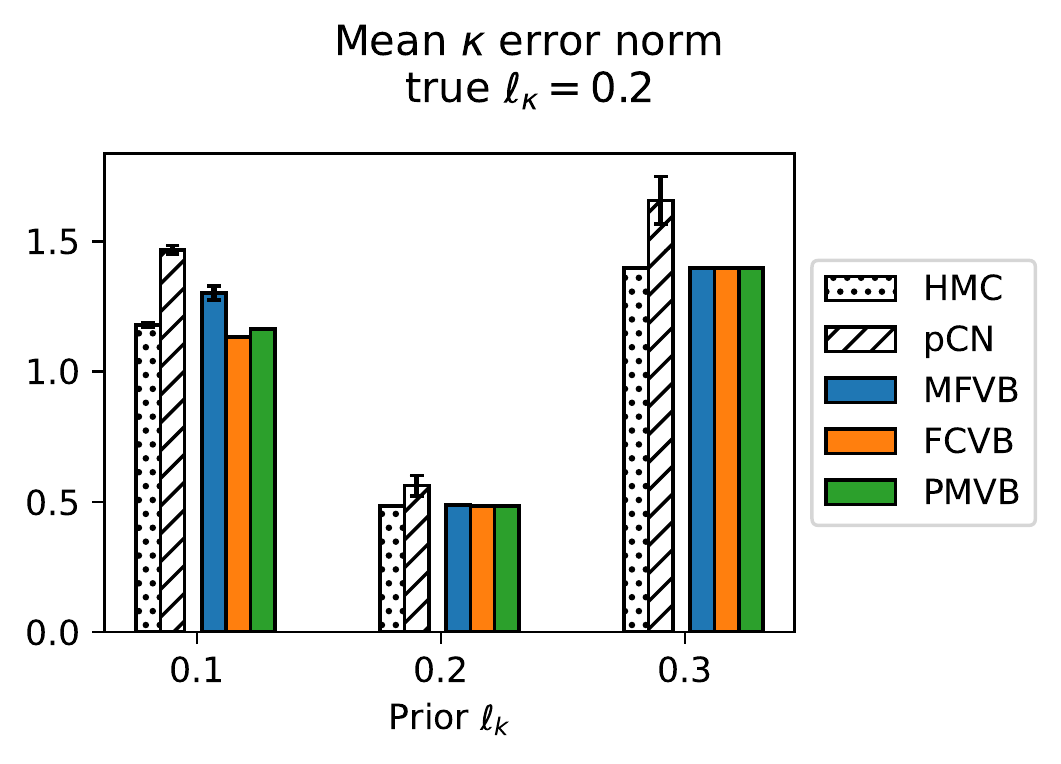}
    \includegraphics[width=0.485\textwidth]{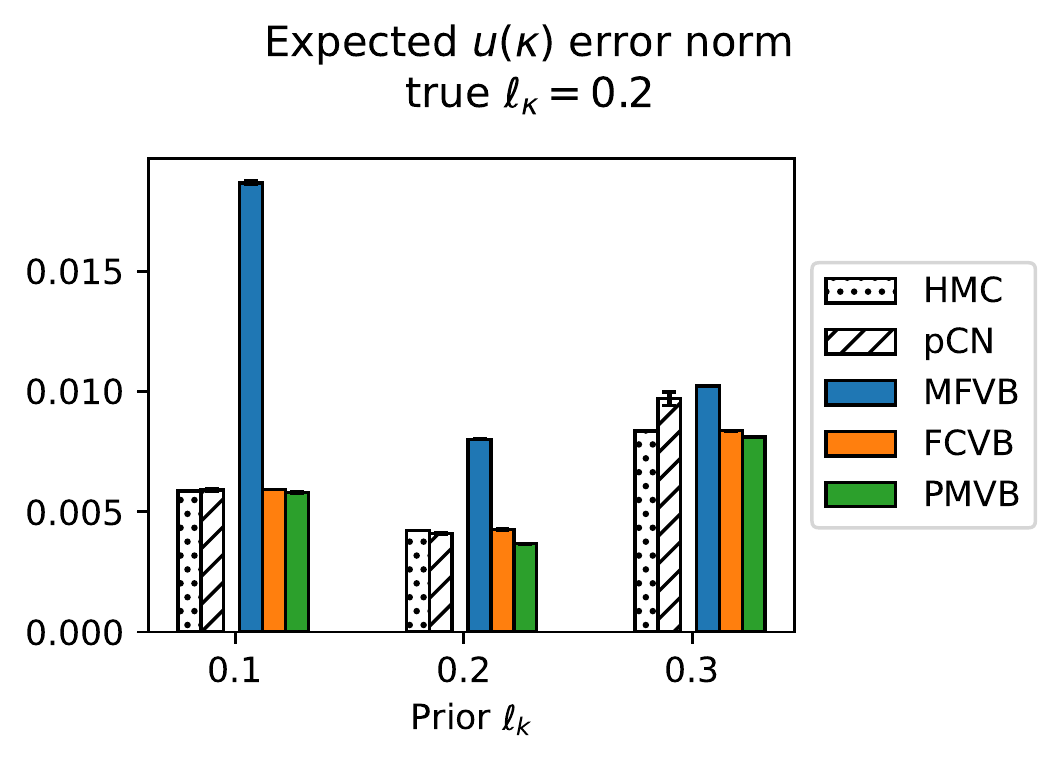}
    \caption{Mean $\kbf$ error norm for the Poisson 2D problem (left), as defined in~\eqref{eq:computations_mean_kappa_error}, and expected solution error norm (right), as defined in~\eqref{eq:computations_expected_solution_error}. Both quantities are estimated using 10,000 samples from the inferred posterior distribution of $\kbf$. Quantitatively, the sampling methods (HMC and pCN) and VB produce comparable results in both metrics, except MFVB parametrisation which captures the mean of $\kbf$ well (as demonstrated in the mean $\kbf$ error norm), but fails to account for the uncertainty as manifested in high error norm in the solution space. For a qualitative comparison, see Fig.~\ref{fig:poisson2d_main_results:1}--\ref{fig:poisson2d_main_results:3}.}
    \label{fig:poisson2d_error_score}
\end{figure}

Firstly, the results in Fig.~\ref{fig:poisson2d_error_score} show that the mean $\kbf$ error of VB methods is very similar to the sampling methods (pCN and HMC). 
Similarly to the 1D case, the expected solution error norm is highest for MFVB estimate, indicating the lack of capturing the possible values of $\kbf$ for which the solutions, $\uu(\kbf)$, are consistent with the observed data. The results also show that both errors are lowest when the prior $\ell_\kappa$ matches the length-scale used to generate the data.

\begin{figure}
\centering
\includegraphics[width=\textwidth]{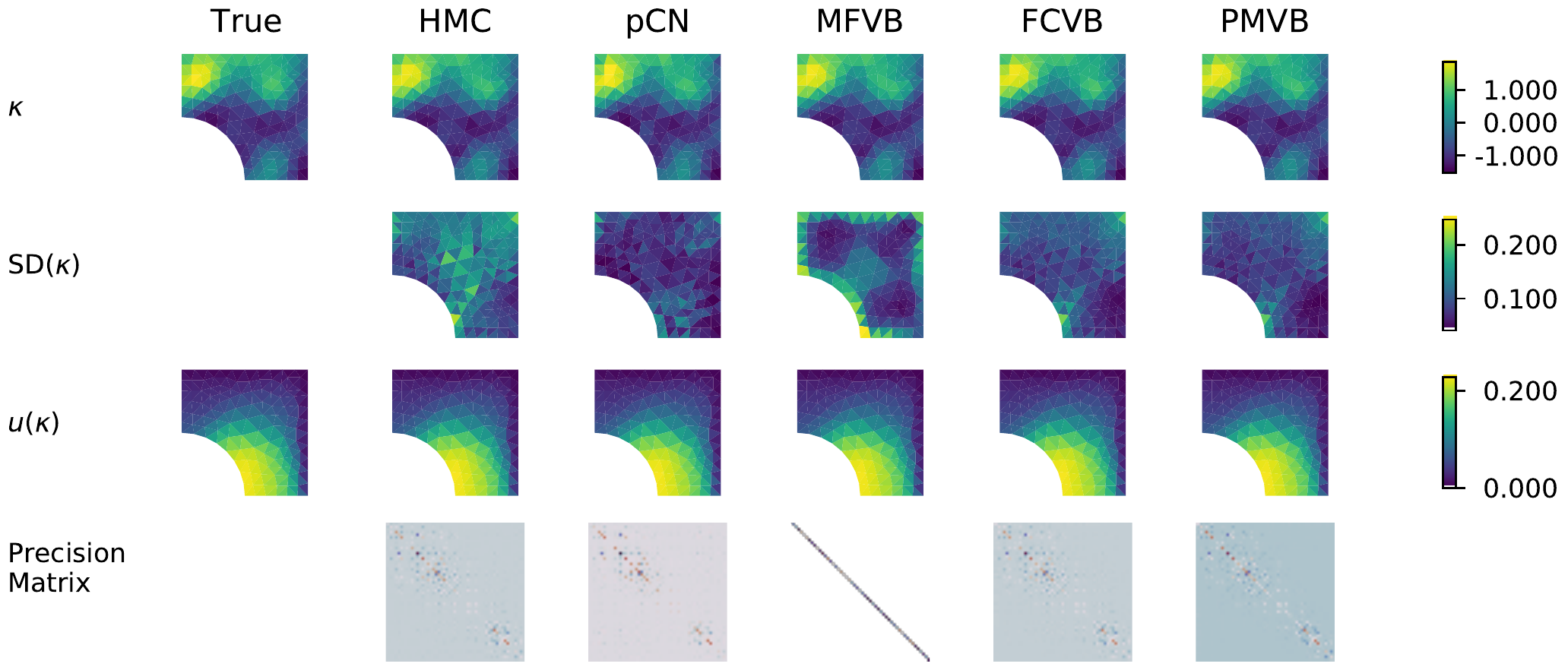} \\
\caption{Posterior mean and standard deviation for $\kbf$ and the corresponding $\uu$ for 2D Poisson example with prior length-scale $\ell_{\kappa} = 0.1$. The bottom row shows the structure of the precision matrix for each inference scheme.}
\label{fig:poisson2d_main_results:1}
\end{figure}

\begin{figure}
\centering
\includegraphics[width=\textwidth]{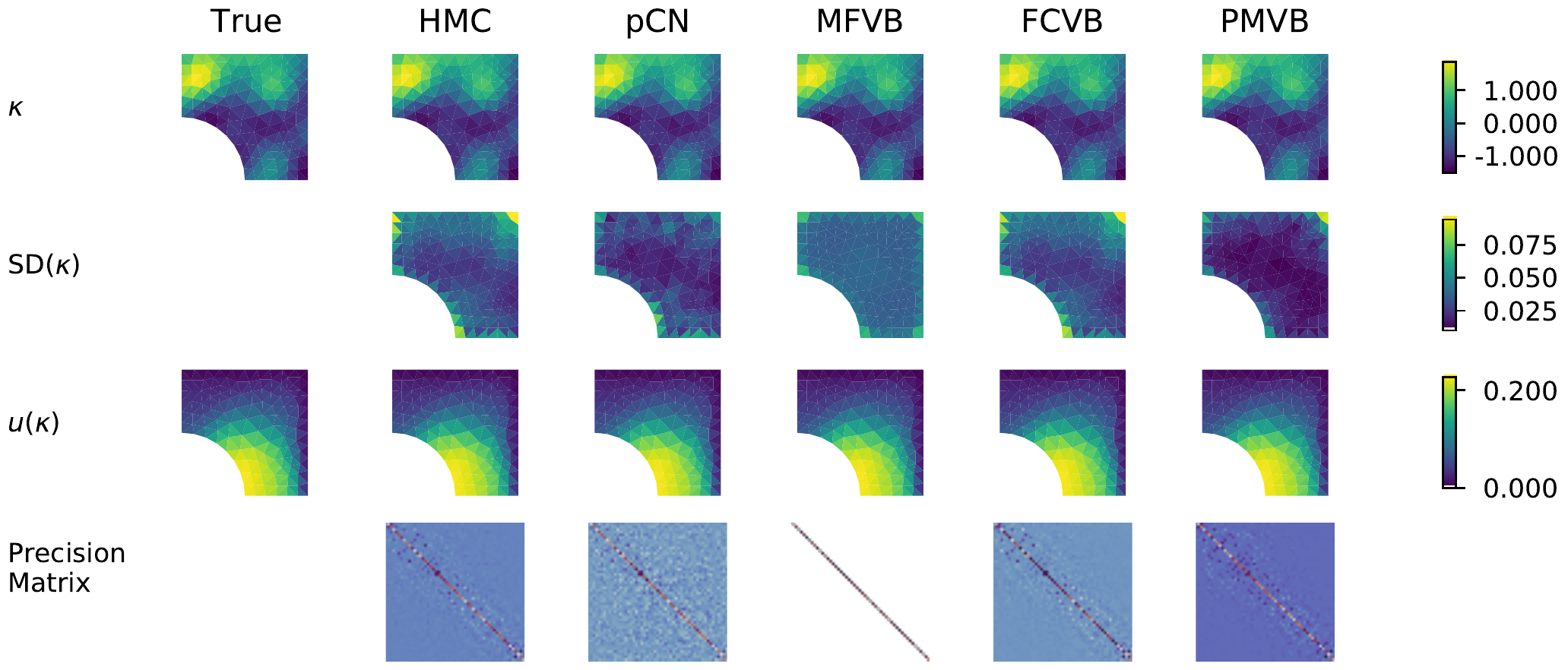}\\
\caption{Posterior mean and standard deviation for $\kbf$ and the corresponding $\uu$ for 2D Poisson example with prior length-scale $\ell_{\kappa} = 0.2$. The bottom row shows the structure of the precision matrix for each inference scheme.}
\label{fig:poisson2d_main_results:2}
\end{figure}

\begin{figure}
\centering
\includegraphics[width=\textwidth]{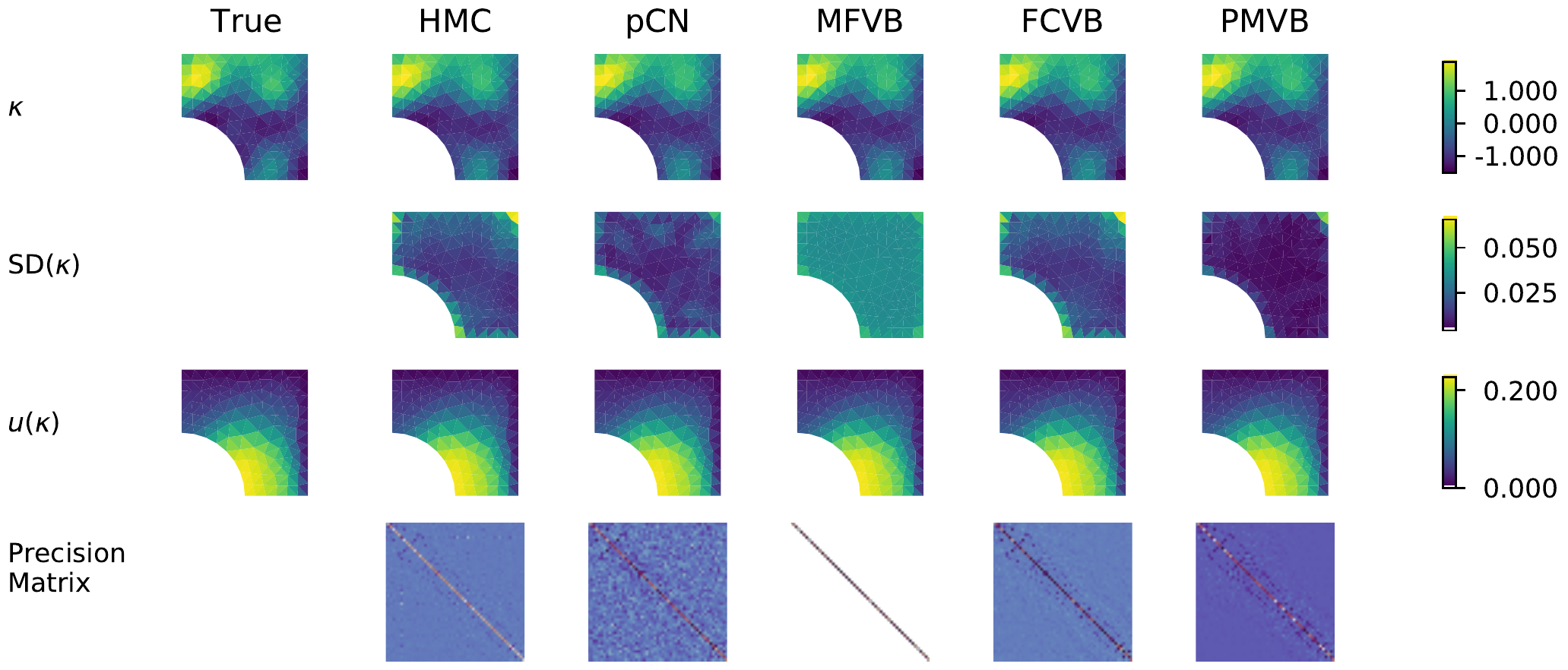}\\
\caption{Posterior mean and standard deviation for $\kbf$ and the corresponding $\uu$ for 2D Poisson example with prior length-scale $\ell_{\kappa} = 0.3$. The bottom row shows the structure of the precision matrix for each inference scheme.}
\label{fig:poisson2d_main_results:3}
\end{figure}

Fig.~\ref{fig:poisson2d_main_results:1}-\ref{fig:poisson2d_main_results:3} show the results for the posterior mean and the standard deviation of $\kbf$, the solution $\uu(\kbf)$ corresponding to the mean of the posterior.  We consider three configurations with prior length-scale $\ell_{\kappa} \in \{0.1, 0.2, 0.3\}$, where the length-scale used to generate the data is $\ell_{\kappa}=0.2$. In all cases, the estimates of the posterior mean of $\kbf$ and the corresponding solutions $\uu$ are very close to the true values. Similarly to the 1D case discussed in Sec.~\ref{sec:1d-poisson}, the variance estimates between HMC and FCVB are consistent, especially for longer prior length-scales. There seems to be a discrepancy between the estimates obtained using MFVB and those obtained by other methods. The estimates obtained using precision-matrix parametrisation are qualitatively very close to the FCVB and MCMC estimates.

The bottom rows of Fig.~\ref{fig:poisson2d_main_results:1}-\ref{fig:poisson2d_main_results:3} show the precision matrices for the inferred posterior distributions. 
As in the one-dimensional examples, the precision matrices of FCVB and PMVB capture similar dependence structure as the one obtained using HMC, implying that PMVB closes the gap between the over-simplified MFVB and the full covariance VB in terms of the captured dependence relationships while retaining a sparse structure.


\begin{figure}
    \centering
    \includegraphics[width=.8\textwidth]{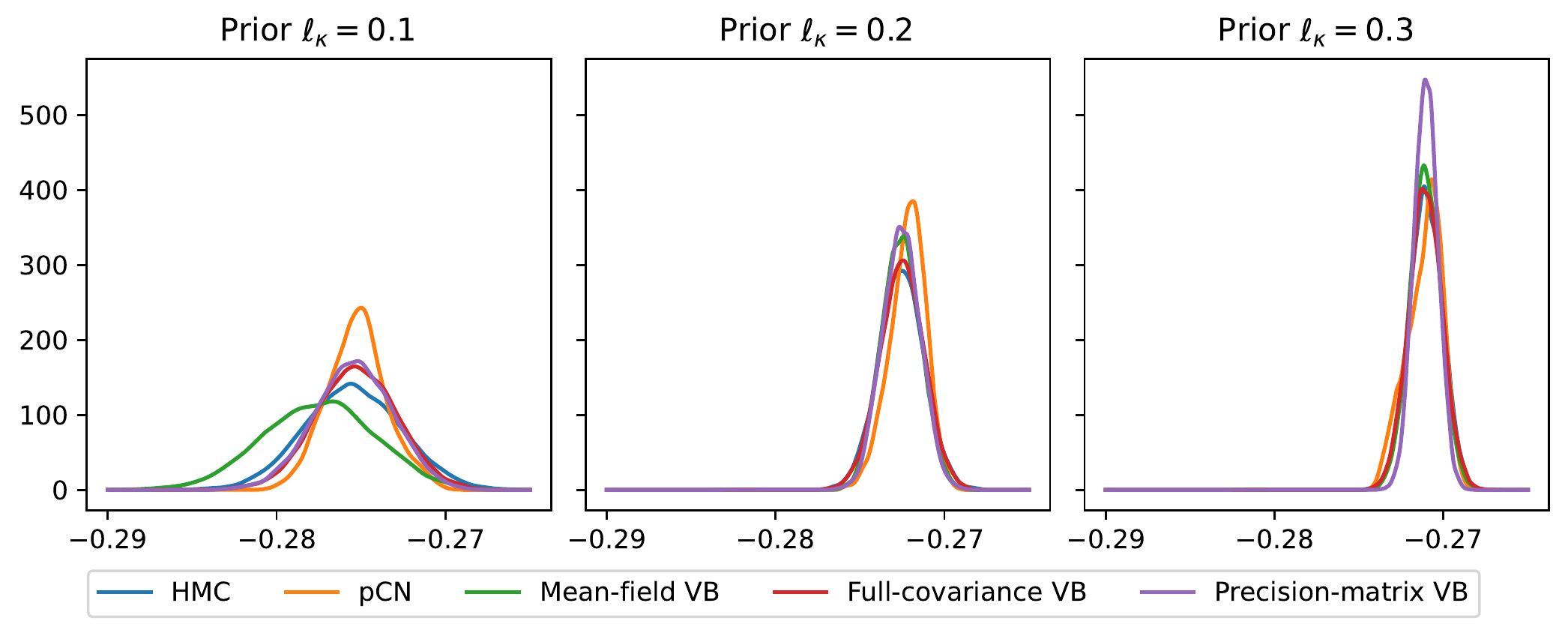}
    \caption{Log of the total flux computed along the right boundary ($x=1$). For PMVB, the precision matrix is parametrised using the second-order neighbourhood structure, as shown in~Fig.~\ref{fig:adjacency_precision_parametrisation}.}
    \label{fig:poisson-2d-qoi-plot}
\end{figure}

For the quantity of interest, we compute the log of the total flux along the right boundary of the domain ($x=1$), and the results are shown in~Fig.~\ref{fig:poisson-2d-qoi-plot}. Unlike the 1D case, the posterior estimates of the boundary flux are approximately the same for all the considered methods, except for the mean-field estimate when prior $\ell_{\kappa}=0.1$, where the MFVB estimate is biased as compared to the other methods.

The empirical computational cost for these experiments is given in Table~\ref{tab:poisson_2d_run_times}. For the HMC experiments, we obtained 250,000 samples, out of which the first 125,000 were used to calibrate the sampling scheme and discarded afterwards. The timing results show that HMC takes an order of magnitude longer than variational Bayes, with some variation that depends on the parametrisation.
\begin{table}
    \centering
    \begin{tabular}{llrrrrr}
    \toprule
    \multirow{2}{*}{true $\ell_\kappa$} & \multirow{2}{*}{prior $\ell_\kappa$} & \multicolumn{5}{c}{Time (hours)} \\
     \cmidrule{3-7}
        &     & \multicolumn{2}{c}{HMC}            &         MFVB &         FCVB   & PMVB \\
    \cmidrule{1-2} \cmidrule{3-7}
0.1 & 0.1 & 240.6 & (930--11200)   &  6.4 & 29.6 & 28.1 \\
    & 0.2 & 295.5 & (1537--11067)  &  6.6 & 32.6 & 28.9 \\
    & 0.3 & 242.0 & (1057--6068)   &  7.3 & 27.3 & 30.6 \\
0.2 & 0.1 & 242.7 & (1102--18235)  &  6.2 & 34.3 & 27.2 \\
    & 0.2 & 264.3 & (1304--9848)   &  7.4 & 33.7 & 34.0 \\
    & 0.3 & 221.9 & (1192--6356)   &  7.8 & 31.3 & 34.0 \\
    \bottomrule
    \end{tabular}
    \caption{Run-times for different inference schemes in seconds. The number of Monte Carlo samples is $N_{\text{SVI}}=5$ for all MFVB, FCVB, and PMVB. The column for HMC includes the range of effective sample sizes (ESS) across different components of $\kbf$.}
    \label{tab:poisson_2d_run_times}
\end{table}

\subsection{Inverse Problem Benchmark}
We evaluate the effectiveness of VB methods on a recently proposed benchmark for Bayesian inverse problems \citep{aristoffBenchmarkBayesianInversion2021}. The benchmark aims to provide a test case that reflects practical applications, but at the same time is easy to replicate. Like above, the test case is a Poisson inverse problem where the task is to recover log-diffusion, $\kappa$, from a finite set of noisy observations. The problem domain is a unit square, the forcing function $f(\vx)=10$ is constant throughout the domain, and the solution of the PDE is imposed to be zero on all four boundaries.

\begin{figure}
    \centering
    \includegraphics[width=.9\textwidth]{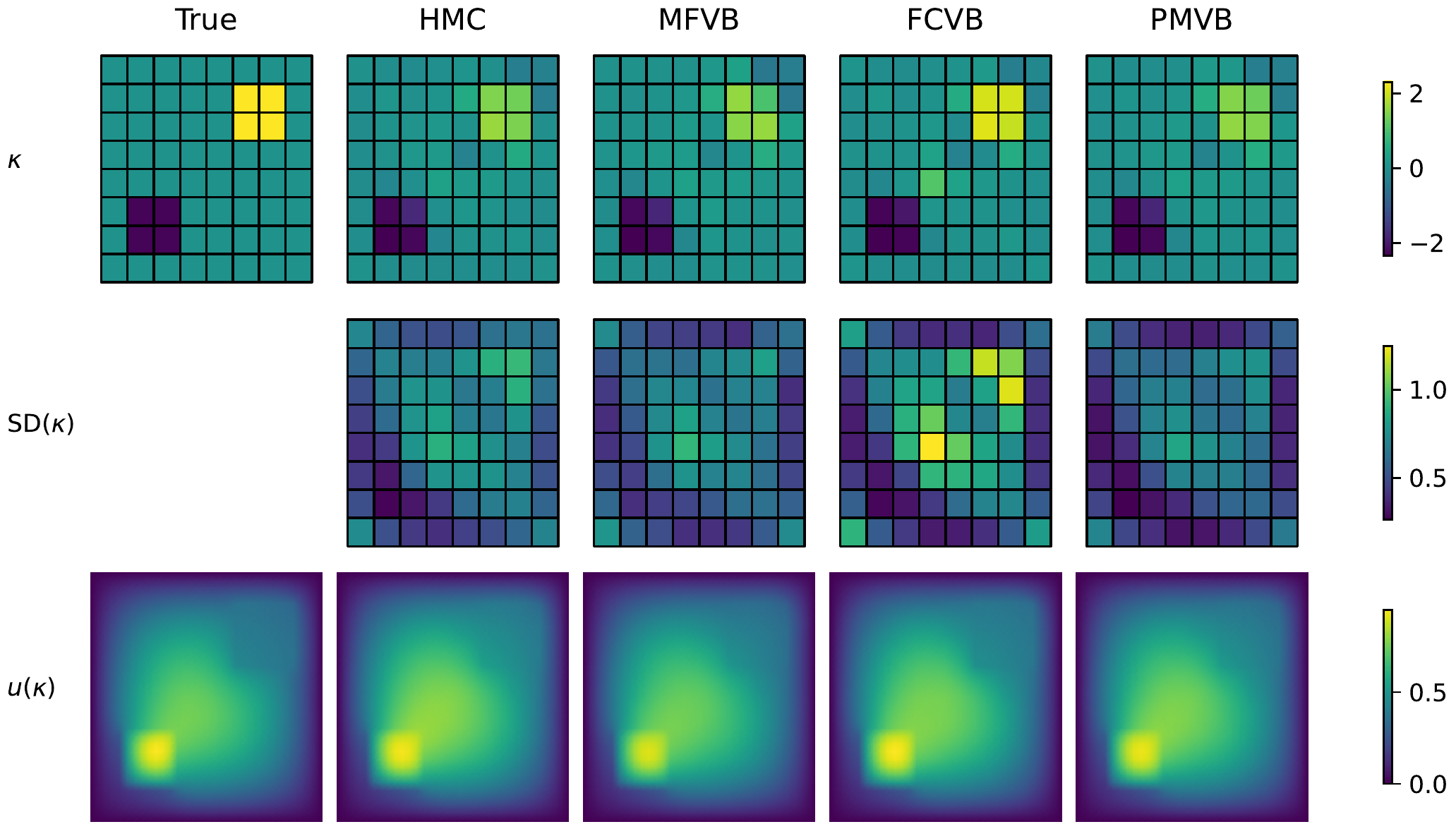}
    \caption{Posterior mean and standard deviation for $\kbf$ and the corresponding $\uu$ for the benchmark example with independent prior for each coefficient of $\kbf$: $\kappa_i \sim \mathcal{N}(0, 1)$.}
    \label{fig:benchmark_results_plot}
\end{figure}

The benchmark discretises $\kappa$ using 64 quadrilateral elements, such that $\kappa$ is constant for each individual element as shown in Fig.~\ref{fig:benchmark_results_plot}. The forward solution of the PDE is obtained after discretising $u$ using $32 \times 32$ bilinear quadrilateral elements. The locations where the solution is observed are placed on a uniform grid of 169 points ($13 \times 13$). The measurements are corrupted by the Gaussian noise with standard deviation $\sigma_\y=0.05$. The authors of the benchmark provide the measurements as well as the true log-diffusion coefficient $\kappa$ which generated the observations. The true log-diffusion coefficient, shown in Fig.~\ref{fig:benchmark_results_plot},  is zero throughout the domain, except two regions, where the value is $\log(10)$ and $\log(0.1)$. It is these two jumps that make it a non-trivial test case. 

Unlike in the previous examples, we place a prior on $\kappa$ which does not induce any spatial correlation between any of the $\kbf$ coefficients. The role of the prior is to express our belief about the ranges of the coefficients, rather than any dependencies. Although authors place $\mathcal{N}(\mu=4, \sigma^2=4)$ for each component of $\boldsymbol{\kappa}$ independently, we choose $\mathcal{N}(\mu=0, \sigma^2=1)$ as most of the coefficients of the true $\kbf$ are at the baseline level equal to zero, and the fact that the $\kbf$ corresponds to the diffusion parameter on the log-scale, a priori we do not expect such high variance.

We performed the inference using HMC, MFVB, FCVB, and PMVB. The means and standard deviations of inferred log-diffusion coefficients, together with the PDE solutions corresponding to the inferred means, are shown in Fig.~\ref{fig:benchmark_results_plot}. The results suggest that the mean estimates of all three methods do capture the jumps and the overall structure of $\kbf$. Specifically, the FCVB estimate of the mean of $\kbf$ is closest to the true value. As for uncertainty quantification, the MFVB and PMVB estimates are closer to the HMC estimate (our assumed ground truth for the uncertainty) than the FCVB estimate. The FCVB estimate seems to overestimate the uncertainty at a few locations. This is potentially due to being stuck in a local optimum during the optimisation procedure, which for FCVB involves high-dimensional exploration.

\subsection{Multimodal Poisson 1D}

One of the advantages of VB is the flexibility of the choice of the trial distribution. To illustrate this, we consider the Poisson equation on the domain~$\Omega=(0, \,1)$ given by
\begin{equation}
    \exp(\kappa) \nabla^2 u(x) = 2,
\end{equation}
where $\exp(\kappa)$ is the conductivity, and the Dirichlet boundary conditions are $u(0) = 0$ and $u(1) = u_R$.

\begin{figure}
    \centering
    \includegraphics[width=0.8\textwidth]{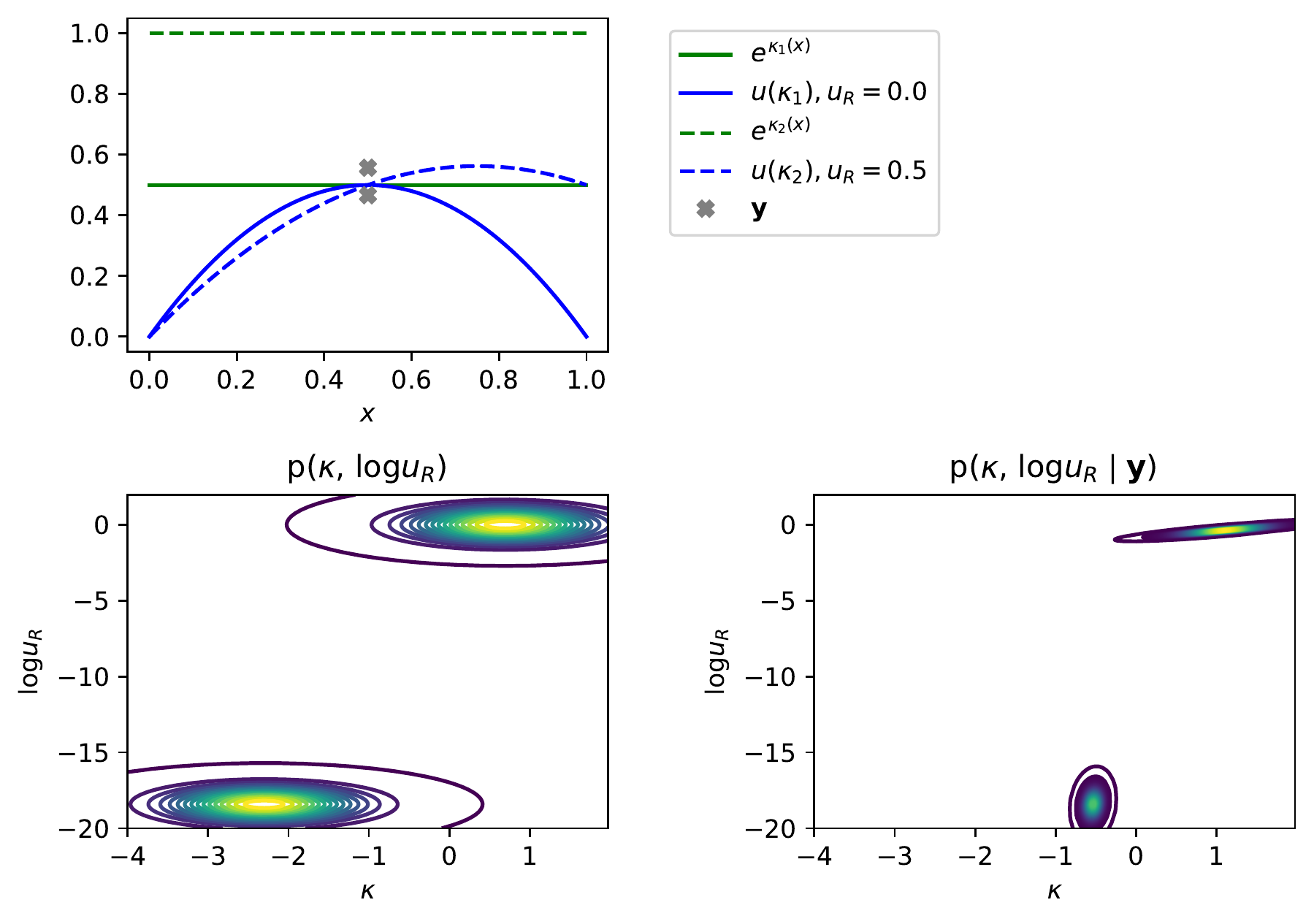}
    \caption{Multimodal Poisson 1D. \textit{Top}: the solution, $u$ (blue) is shown for two different conductivities and boundary conditions on the right. Two measurements that were taken at the centre of the domain are marked as crosses. \textit{Bottom left}: joint prior distribution for conductivity and the Dirichlet boundary condition on the right at $x=1$. \textit{Bottom right}: the posterior distribution inferred from the prior distribution and the two measurements (shown in the top panel).}
    \label{fig:poisson-1d-bimodal-example}
\end{figure}

We are interested in inferring the constant conductivity, $\exp(\kappa)$, and the right boundary condition $u_R$, having obtained multiple measurements of $u(x)$ at $x=0.5$. We show the solution of this problem in the top part of Fig.~\ref{fig:poisson-1d-bimodal-example} where we consider two different combinations of $\kappa$ and $u_R$ that result in the same solution $u(0.5)$. This implies that there are multiple combinations of the two unknown parameters that result in the same solution at the observation point, making the inference problem ill-posed.

To proceed, we place a prior distribution which is motivated by the domain knowledge: if conductivity is high, so will be the solution $u_R$ at the right boundary. A mixture model provides a convenient way of encoding this prior information in a probability distribution. Specifically, we place the following prior consisting of two bivariate Gaussian distributions on the log of conductivity and the log of boundary condition $u_R$:
\begin{align}
    \begin{pmatrix}
       \kappa \\
      \log u_R
    \end{pmatrix} & = \frac{1}{2} \mathcal{N}
    \begin{pmatrix}
        \begin{pmatrix}
          \log 0.1 \\
          \log 1\times 10^{-8}
        \end{pmatrix},
        0.5 
        \begin{pmatrix}
          1 & 0 \\
          0 & 1 \\
        \end{pmatrix}
    \end{pmatrix} 
    + \frac{1}{2} \mathcal{N}
    \begin{pmatrix}
        \begin{pmatrix}
          \log 2.0 \\
          \log 1.0
        \end{pmatrix},
        0.5 
        \begin{pmatrix}
          1 & 0 \\
          0 & 1 \\
        \end{pmatrix}
    \end{pmatrix} 
  \end{align}
The contour plot of this prior is shown in the bottom-left part of Fig.~\ref{fig:poisson-1d-bimodal-example}.

Assuming a Gaussian measurement noise with $\sigma_y = 0.05$, we take two samples of the temperature at the observation point. Following the variational Bayes approach, we restrict the family of trial distributions to be an equally weighted mixture of bivariate Gaussian distributions, each with its own mean and covariance matrix, parametrised by the Cholesky factor. As there is no closed-form expression for the KL divergence between the prior and members of the family of trial distributions, we estimate the KL divergence term in the ELBO using Monte Carlo sampling. As shown in the bottom right panel of Fig.~\ref{fig:poisson-1d-bimodal-example}, the resulting posterior distribution is bimodal. The distribution is consistent with the physical intuition which we expressed in the prior.

This illustrative example shows that when a proposed model exhibits multi-modality, the flexibility of the variational Bayes methodology allows for specifying a family of trial distributions that can capture that property.

\section{Conclusions}\label{sec:conclusion}
In this paper, we have presented the variational inference framework for Bayesian inverse problems and investigated its efficacy on problems based on elliptic PDEs. Computationally, variational Bayes offers a tractable alternative to the intractable MCMC methods, and provides consistent mean and uncertainty estimates on the problems inspired by questions in computational mechanics. 
VB recasts the integration problem associated with Bayesian inference into an optimisation problem.
As such, it is naturally integrated with existing FEM solvers, using the gradient calculations from the FEM solvers to optimise the ELBO in VB. Furthermore, the geometry of the problem encoded in the FEM mesh is utilised through the use of a sparse precision matrix that defines the conditional independence structure of the problem.
Our results on the 1D and 2D Poisson problems support the claims of accuracy and scalability of VB. We note that the inferred variance is important in uncertainty quantification with a probabilistic forward model (for a different load case). 

More specifically, our results show that
\begin{itemize}
    \item the mean of the variational posterior provides an accurate point estimate irrespective of the choice of the parametrisation of the covariance structure,
    \item the variational approximation with a full-covariance or precision matrix structure adequately estimates posterior uncertainty when compared to HMC and pCN which are known to be asymptotically correct,
    \item parametrising the multivariate Gaussian distribution using a sparse precision matrix provides a way to balance the trade-off between computational complexity and the ability to capture dependencies in the posterior distribution,
    \item variational Bayes provides a good estimate for the mean and the variance of the posterior distribution in a time that is an order of magnitude faster than HMC or pCN,
    \item the multivariate Gaussian variational family is flexible enough to capture the true posterior distribution with high accuracy,
    \item the VB estimates may be used effectively in downstream tasks to estimate various quantities of interest, and
    \item variational Bayes method is flexible enough to model multimodal posteriors, as illustrated on the steady-state heat equation.
\end{itemize}

Our work may be extended in a number of natural ways that allows for greater adaptivity to the specific problems encountered in applications and integration within existing frameworks. Firstly, taking advantage of fast implementations of sparse linear algebra routines would further improve the scalability of VB with the structured precision matrix, as proposed in our work. Secondly, casting the inverse problem in a multi-level setting and taking advantage of low-dimensional projections has potential to further improve computational efficiency~\citep{nagelUnifiedFrameworkMultilevel2016a, ghattasLearningPhysicsbasedModels2021}.
Thirdly, the results provided in this paper use standard off-the-shelf optimisation routines; further computational improvements may be achieved using customised algorithms. 
As a further extension, in some applications it may be informative to consider the uncertainty in the forcing function so that the forward mapping is stochastic, as discussed in~\citep{girolamiStatisticalFiniteElement2021}.
Finally, one of the aims of our work is to take advantage of the advances in Bayesian inference and adapt the novel algorithms to inverse problems in computational mechanics. As such, any further developments in VB as applied to machine learning and computational statistics problems may be directly applied using the framework proposed in this paper.

\section{Implementation}
\label{sec:bip-implementation}
Codes for performing all forms of variational Bayes inference presented in this paper are available on Github at \texttt{\url{https://github.com/jp2011/bip-pde-vi}}. The user must provide their own PDE solver which accepts $\kbf$ as input parameter and computes $\log p(\boldsymbol{y} \mid \kbf)$, together with its gradient with respect to $\kbf$.

\section*{Acknowledgements}
We would like to thank William Baldwin for providing us the necessary background for and the implementation of the bimodal example; Garoe Dorta for his help with Tensorflow debugging. JP, IK, and MG were supported by the EPSRC grant EP/P020720/2 for Inference, COmputation and Numerics for Insights into Cities (ICONIC, \url{https://iconicmath.org/}). JP was also supported by EPSRC (EP/L015129/1). MG was supported by EPSRC grants EP/T000414/1, EP/R018413/2, EP/R034710/1, EP/R004889/1, and a Royal Academy of Engineering Research Chair in Data Centric Engineering.

\newpage
\appendix

\section{Short length-scale results} \label{sec:short_lengthscale_results}

\begin{figure}[h!]
\centering
\includegraphics[width=.485\textwidth]{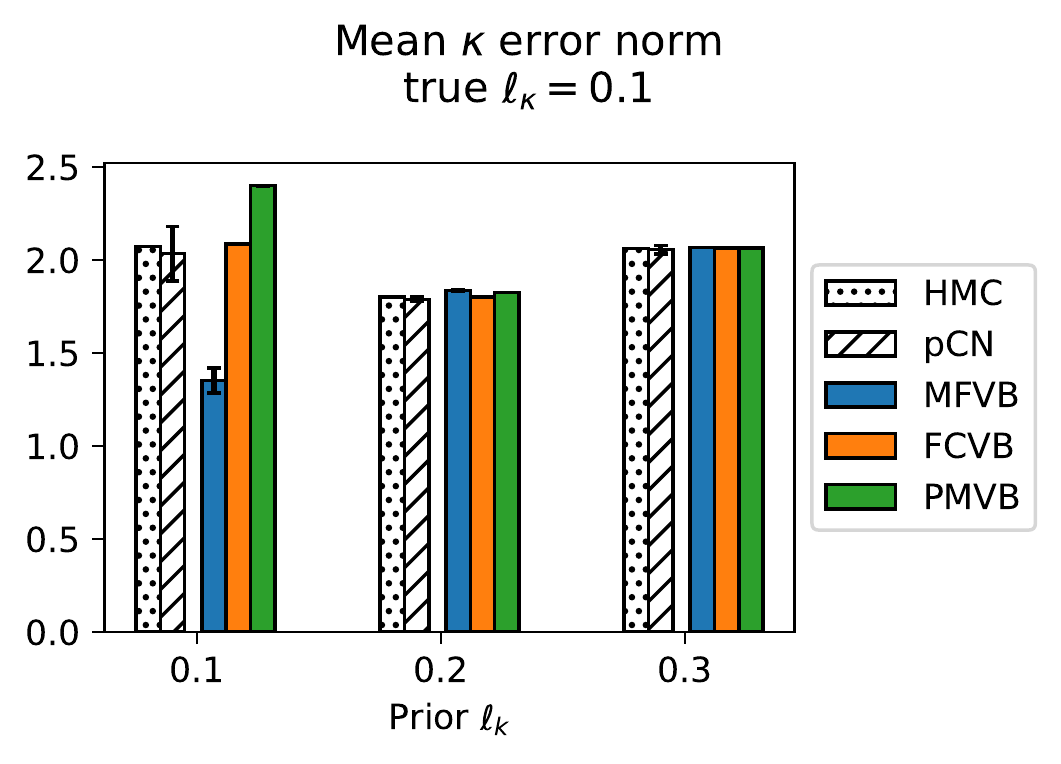} 
\includegraphics[width=.485\textwidth]{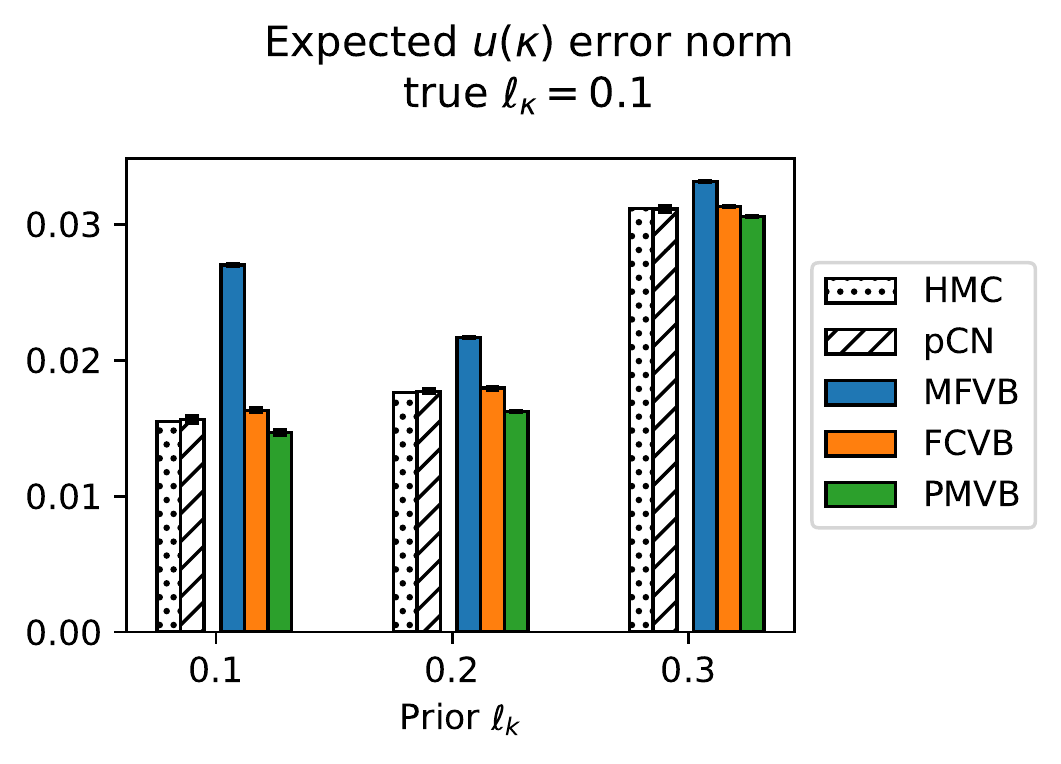} 
\caption{Mean $\kbf$ error norm for the Poisson 1D problem (left), as defined in~\eqref{eq:computations_mean_kappa_error}, and expected solution error norm (right), as defined in~\eqref{eq:computations_expected_solution_error}. Both quantities are estimated using 10,000 samples from the inferred posterior distribution of $\kbf$. Quantitatively, the sampling methods (HMC and pCN) and VB produce comparable results in both metrics, except MFVB parametrisation which captures the mean of $\kbf$ very well, but fails to account for the uncertainty as manifested in high error norm in the solution space. For a qualitative comparison, see Fig.~\ref{fig:poisson1d_main} where each row of results corresponds to a different value of the true prior length-scale $\ell_\kappa$.}
\label{fig:appendix_poisson1d_score_and_error_short_lengthscale}
\end{figure}

\begin{figure}
\centering
\includegraphics[width=0.8\textwidth]{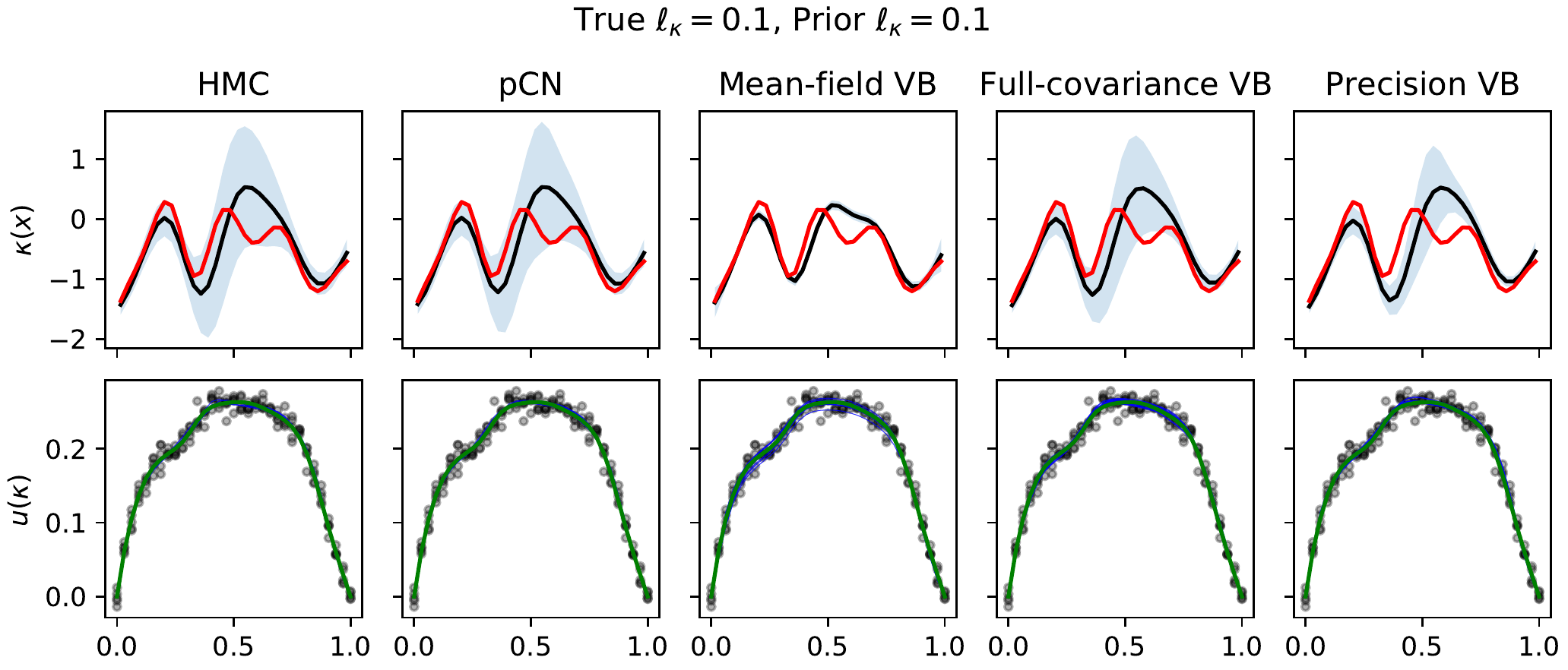}
\\
\vspace{1.2em}
\includegraphics[width=0.8\textwidth]{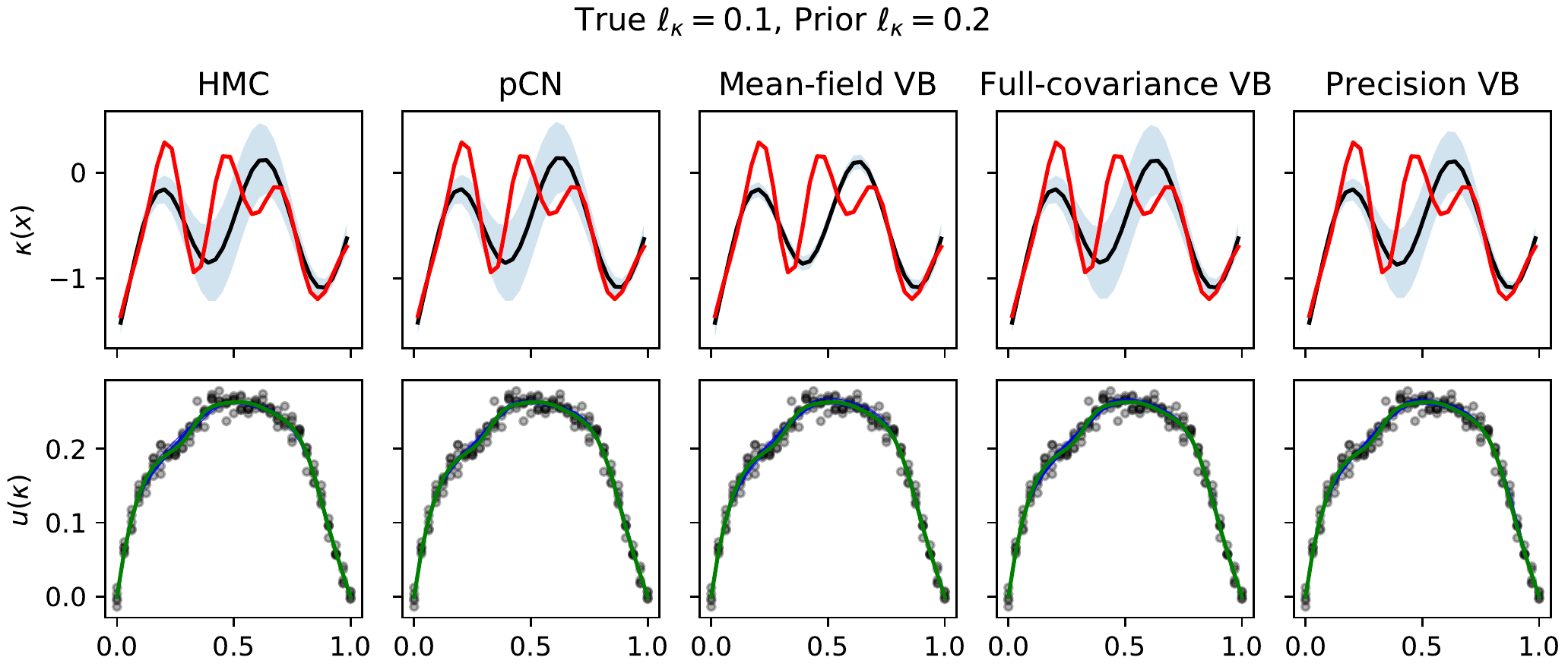}
\\
\vspace{1.2em}
\includegraphics[width=0.8\textwidth]{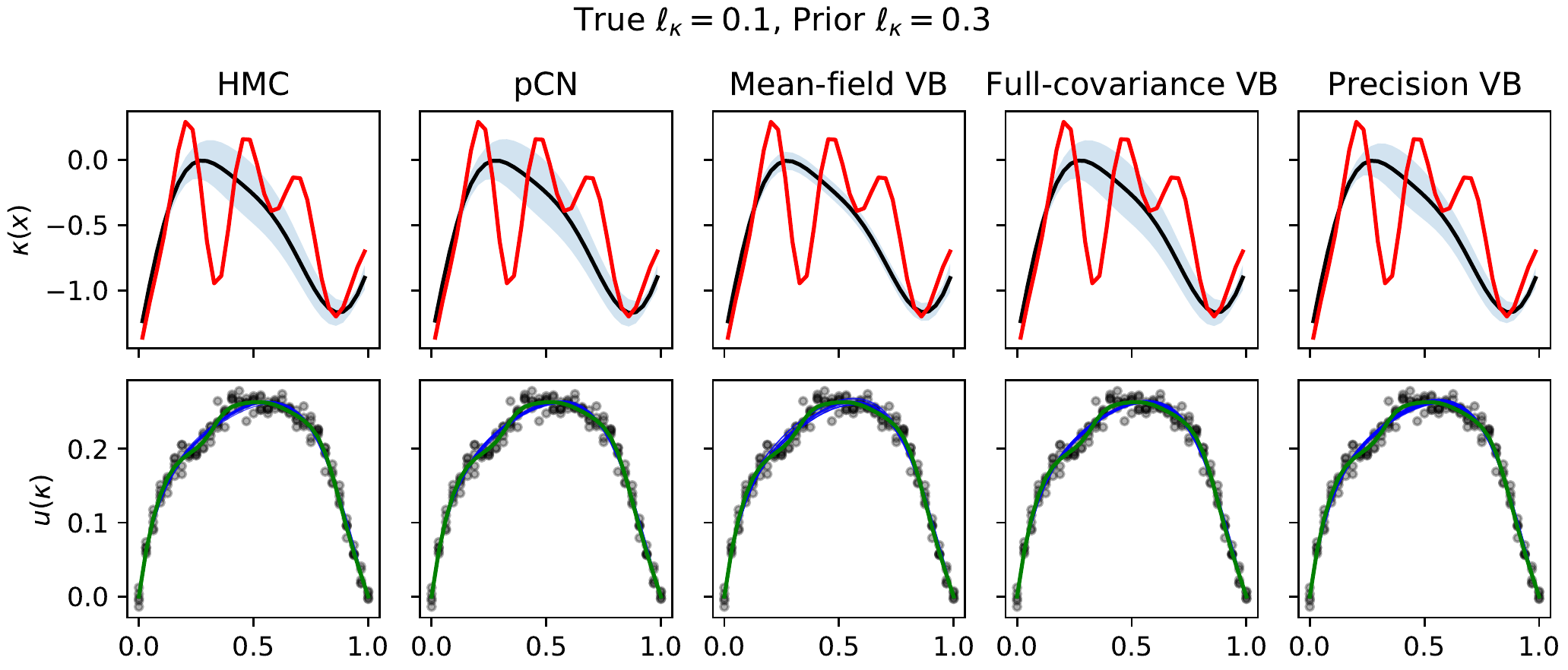}
\\
\includegraphics[width=0.8\textwidth]{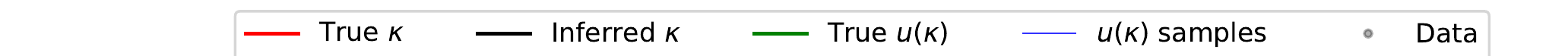}
\caption{Top row in each of the three panels show true values of $\kappa(x)$ (red), posterior means (black) and posterior variances (blue shaded regions) for HMC and VB variants for different values of prior length-scales $\ell_{\kappa}$. The bottom rows show the data (black), true solution $\uu$ (green), solutions for different samples of $\kappa$ (blue). For the PMVB estimate, the bandwidth is set to $10$.}
\label{fig:poisson1d_appendix_short_lengthscale}
\end{figure}

\begin{figure}
    \centering
    \includegraphics[width=\textwidth]{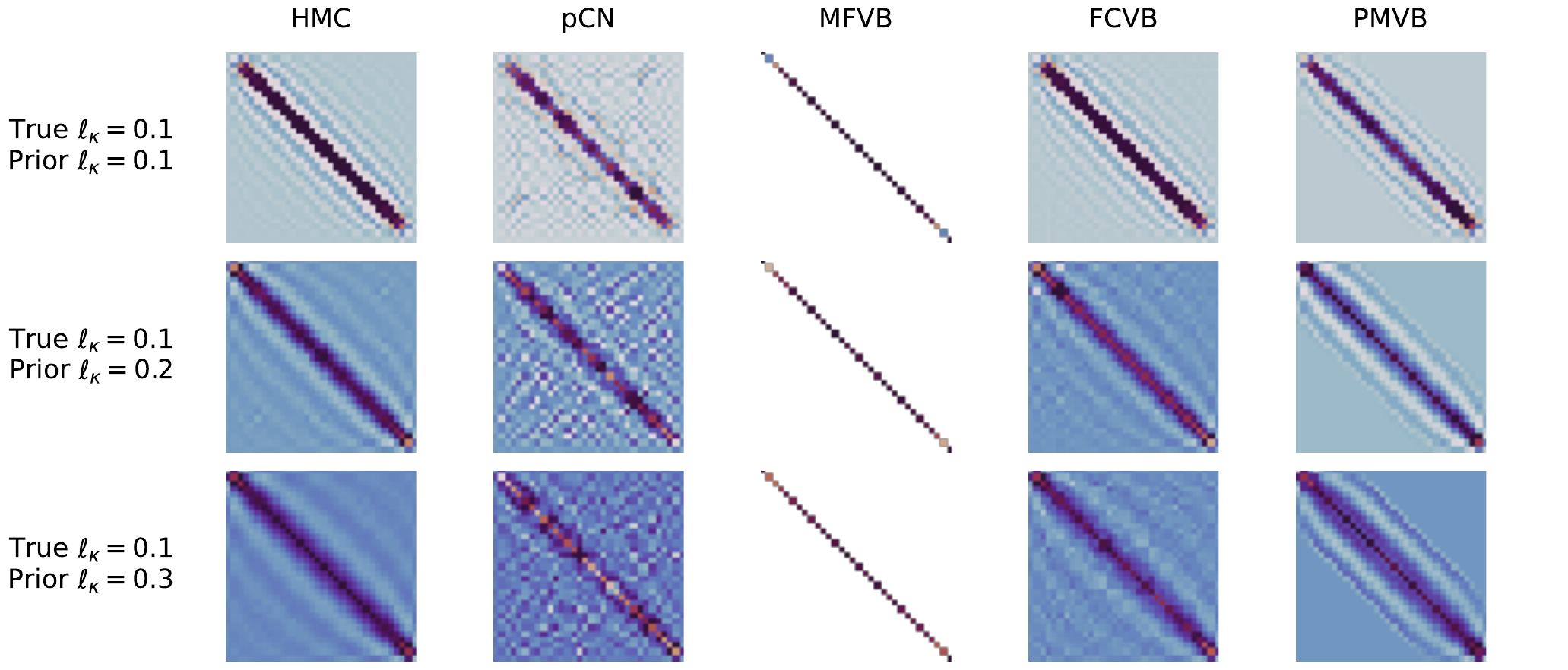}
    \caption{Precision matrices for each of the considered methods, where true $\ell_\kappa = 0.1$ and each row corresponds to a different value of prior $\ell_\kappa$.}
    \label{fig:poisson1d-precision-matrix-sparsity-structure-truel-len-0_1}
\end{figure}

\section{Variational Inference}
\subsection{Reparametrisation trick}
\label{sec:appendix_reparametrisationa_trick}

Reparametrisation trick allows computing the gradients of quantities derived from samples from a probability distribution with respect to the parameters $\rvphi$ of that probability distribution. This holds for probability distributions where samples can be obtained by a deterministic mapping, parametrised by $\rvphi$, of other random variables. 

Let $\epsilon$ be a set of random variables. We assume that samples of $\kbf \sim q(\kbf ; \rvphi)$ are given by a deterministic mapping
\begin{equation}
\kbf = t(\rvphi, \epsilon).
\end{equation}
The KL divergence between approximating distribution $q(\kbf)$ and the prior $p(\kbf)$ is often available in closed form and so are its gradients with respect to $\rvphi$. To estimate the gradients of the Monte Carlo estimate of the log-likelihood of the data,
\begin{equation}
\E_q\big[ \log p(\vy \mid \kbf) \big] \approx
  N_{\text{SVI}}^{-1}\sum_{i=1}^{N_{\text{SVI}}}\log p(\vy \mid \kbf^{(i)}),
\end{equation}
we can use the chain rule of differentiation to obtain
\begin{align}
\nabla_{\rvphi} N_{\text{SVI}}^{-1}\sum_{i=1}^{N_{\text{SVI}}}\log p(\vy \mid \kbf^{(i)})   
& = N_{\text{SVI}}^{-1}\sum_{i=1}^{N_{\text{SVI}}} \nabla_{\kbf} \log p(\vy \mid \kbf^{(i)}) \cdot \nabla_{\rvphi} t(\rvphi, \epsilon^{(i)}).
\end{align}

\section{Markov Chain Monte Carlo}

\subsection{Pre-Conditioned Crank-Nicholson Scheme}\label{sec:pcn_algo_section}
We consider the pre-conditioned Crank-Nicholson scheme proposed by~\citet{cotterMCMCMethodsFunctions2013}. We summarise the procedure in Algorithm~\ref{alg:pcn}.
\begin{algorithm}
	\DontPrintSemicolon
	\KwIn{ $\Phi(\kappa, \vy) = - \log p(\vy \mid \kappa)$: likelihood of the data, $\mu_0(\kappa)$: prior measures, $\beta$: corresponds to the amount of innovation in the proposal. If the value is small, there is little innovation and the proposed sample will be close to the previous sample.}
	\KwOut{A list of samples from $\mu^{\y}(\kappa)$.}
	\For{$t \gets 1, 2, \dots$}{
		Sample $\xi^{(t)} \sim \mu_0(\kappa)$ \;
		$v^{(t)} \gets \sqrt{\left(1-\beta^{2}\right)} \kappa^{(t)} +\beta \xi^{(t)}$ \;
    	$\kappa^{(t+1)} \gets \begin{cases}
    		v^{(t)}      & \text{with probability} \min \bigg(1, \exp \big( \Phi(\kappa^{(t)}; \vy) - \Phi(v^{(t)}; \vy) \big)\bigg) \\
    		\kappa^{(t)} & \text{otherwise}
    	\end{cases}$ \;
	}
	\Return{$[\kappa^{(1)}, \kappa^{(2)}, \dots ]$}\;
	\caption{{\sc pre-conditioned Crank-Nicholson MCMC}~\citep{cotterMCMCMethodsFunctions2013}}
	\label{alg:pcn}
\end{algorithm}


\subsection{Hamiltonian Monte Carlo}
\label{sec:hmc_algo_section}
Hamiltonian Monte Carlo~\citep{duaneHybridMonteCarlo1987}
is a variant of Metropolis-Hastings~\citep{metropolisEquationStateCalculations1953, hastingsMonteCarloSampling1970} which takes advantage of the gradients of the target distribution in the proposal, allowing for a more rapid exploration of the sample space, even in a high-dimensional target space. For each component $\kbf_i$ of the target space, the scheme adds a `momentum' variable $\rvphi_j$ (note that this is different from $\rvphi$ used in~\ref{sec:appendix_reparametrisationa_trick}). Subsequently, $\kbf$ and $\rvphi$ are updated jointly in a series of updates to propose a new sample $(\kbf^*, \rvphi^*)$ that is then accepted or rejected. 

The proposal is largely driven by the momentum variable. The proposal step starts with drawing a new value of $\rvphi$ from $p(\rvphi)$ which needs to be specified. Then in a series of user-specified steps, $L$, the momentum variable $\rvphi$ is updated based on the gradient of the $\log$ of the target density, and $\kbf$ is moved based on the momentum. Usually, the distribution of the momentum variable is $\mathcal{N}(\boldsymbol{0}, \boldsymbol{M})$, where $\boldsymbol{M}$ is the so called `mass' matrix. A diagonal matrix is often chosen to be able to efficiently sample from the momentum distribution. The full steps of the procedure are given in Algorithm~\ref{alg:hmc}.
\begin{algorithm}
	\DontPrintSemicolon 
	\KwIn{ $p(\kbf \mid \vy)$: unnormalised target density, $p(\rvphi)$: momentum density and its mass matrix $\boldsymbol{M}$, $L$: leapfrog steps, $\epsilon$: scaling factor}
	\KwOut{A list of samples from $p(\kbf \mid \vy)$.}
	\For{$t \gets 1, 2, \dots$}{
		Sample $\rvphi$ from $p(\rvphi)$ \;
		\For{$i \gets 1$ \textbf{to} $L$} {
		    $\kbf^* \gets \kbf^{t-1}$ \;
			$\rvphi \gets \rvphi + \frac{1}{2}\epsilon \frac{d \log p(\kbf^*|\vy)}{d\kbf}$
\;
			$\kbf^* \gets \kbf^{*} + \epsilon M^{-1}\rvphi$ \;
			$\rvphi \gets \rvphi + \frac{1}{2}\epsilon \frac{d \log p(\kbf^*|\vy)}{d\kbf}$ \;
			$r \gets \frac{p(\kbf^* | \y)p(\rvphi^*)}{p(\kbf^{t-1} | \vy)p(\rvphi^{t-1})}$ \;
			$\kbf^{t} \gets \begin{cases}
				\kbf^*     & \text{with probability }\min(r,1) \\
				\kbf^{t-1} & \text{otherwise}
			\end{cases}$ \;
		}
	}
	\Return{$[\kbf^1, \kbf^2, \dots ]$}\;
	\caption{{\sc Hamiltonian Monte Carlo} as presented in \citet{gelmanBayesianDataAnalysis2013}}
	\label{alg:hmc}
\end{algorithm}

The performance of the algorithm can be tuned in three ways: (i) choice of the momentum distribution $p(\phi)$, which in the version above requires specifying the mass matrix, (ii) adjusting the scaling factor of the leapfrog step, $\epsilon$, and (iii) the number of leapfrog steps, $L$. \citet{gelmanBayesianDataAnalysis2013} suggest setting $\epsilon$ and $L$ so that $\epsilon L = 1$. They suggest tuning these so that the acceptance rate is about $65\%$. As for the mass matrix, the authors suggest that it should approximately scale with the inverse covariance matrix of the posterior distribution, $(\text{Cov}(\kbf \mid \vy))^{-1}$. This can be achieved by a pre-run from which the empirical covariance matrix can be computed.

\Urlmuskip=0mu plus 1mu\relax
\bibliographystyle{model3-num-names}
\bibliography{references}
\end{document}

%% file: defs.tex
\DeclareMathOperator*{\argmax}{arg\,max}
\DeclareMathOperator*{\argmin}{arg\,min}

\DeclareMathOperator{\E}{\mathbb{E}}
\DeclareMathOperator{\ELBO}{ELBO}

\DeclareMathOperator{\KLOP}{KL}
\DeclarePairedDelimiterX{\DistanceRelation}[2]{(}{)}{%
  #1\;\delimsize\|\;#2%
}
\newcommand{\kl}{\KLOP\DistanceRelation}

\newcommand{\X}{\mathbf{X}}

\newcommand{\y}{y}
\newcommand{\veta}{\boldsymbol{\eta}}
\newcommand{\vy}{\boldsymbol{y}}

\newcommand{\x}{x}
\newcommand{\vx}{\boldsymbol{x}}
\newcommand{\K}{\mathbf{K}}
\newcommand{\kbf}{\boldsymbol{\kappa}}
\newcommand{\rvphi}{\boldsymbol{\phi}}

\newcommand{\RR}{\mathbb{R}} 
\newcommand{\uu}{\mathbf{u}} 
\newcommand{\f}{\mathbf{f}}

\newcommand{\Abf}{\mathbf{A}}